\pdfoutput=1
\documentclass[fleqn,usenatbib,usedcolumn]{mnras}
 
\usepackage[british]{babel}             
\usepackage{txfonts}                    
\usepackage[T1]{fontenc}                
\usepackage{bm}

\usepackage{mathptmx}

\usepackage{ae,aecompl}

\usepackage{amssymb}
\usepackage{times}
\usepackage{pifont} 
\usepackage{rotating}
\usepackage[toc,page]{appendix}
\usepackage[dvipsnames,usenames]{color}

\newcommand{\astec}{\textsc{ASTEC}}

\newcommand{\adipls}{\textsc{ADIPLS}}

\newcommand{\dP}{$\Delta P$}

\newcommand{\msol}{M$_\odot$}
\newcommand{\rsol}{R$_\odot$}
\newcommand{\lsol}{L$_\odot$}
\newcommand*{\Resize}[2]{\resizebox{#1}{!}{$#2$}}%

\def\note #1]{\noindent{\bf #1]}}

\def\sgnk2{{\rm sgn\left(K^2\right)}}
\def\rmd{{\rm d}}
\def\rmdd{{\rm d}^2}
\def\mN{\mathcal{N} }

\def\mF{\mathcal{F} }

\title[Period-spacing analytical formulations]{Analytical modelling of period spacings across the HR diagram}
\author[M. S. Cunha, P.P. Avelino et al.]{M.~S.~~Cunha,$^{1,2}$
	P.~P.~Avelino,$^{1,2,3}$ 
	J.~Christensen-Dalsgaard,$^{4}$
	 D.~Stello,$^{4,5,6}$ 
	 M.~Vrard,$^{1}$ \newauthor
	 C.~Jiang,$^{7}$ and 
	 B.~Mosser,$^{8}$.\\
$^{1}$Instituto de Astrof\'\i sica e Ci\^encias do Espa\c co, Universidade do
Porto, CAUP, Rua das Estrelas, PT4150-762 Porto, Portugal\\
$^{2}$ School of Physics and Astronomy, University of Birmingham,Birmingham, B15 2TT, United Kingdom\\
$^{3}$ Departamento de F\'\i sica e Astronomia, Faculdade de Ci\^encias, Universidade do Porto, Rua do Campo Alegre 687, PT4169-007 Porto, Portugal\\
$^{4}$Stellar Astrophysics Centre, Department of Physics and Astronomy, Aarhus University, Ny Munkegade 120, DK-8000 Aarhus C, Denmark\\
$^{5}$School of Physics, The University of New South Wales, Sydney NSW 2052, Australia\\
$^{6}$ Sydney Institute for Astronomy (SIfA), School of Physics, University of Sydney, NSW 2006, Australia\\
$^{7}$School of Physics and Astronomy, Sun Yat-Sen University, Guangzhou, 510275, China\\
$^{8}$LESIA, Observatoire de Paris, PSL Research University, CNRS, Sorbonne Universit\'e, Universit\'e Paris Diderot,  92195 Meudon, France 
}

\date{Accepted XXX. Received YYY; in original form ZZZ}

\pubyear{2019}

\begin{document}
\label{firstpage}
\pagerange{\pageref{firstpage}--\pageref{lastpage}}
\maketitle

\begin{abstract}
The characterisation of stellar cores may be accomplished through the modelling of asteroseismic data from stars exhibiting either gravity-mode or mixed-mode pulsations, potentially shedding light on the physical processes responsible for the production, mixing, and segregation of chemical elements.  In this work we validate against model data an analytical expression for the period spacing that will facilitate the inference of the properties of stellar cores, including the detection and characterisation of buoyancy glitches (strong chemical gradients). This asymptotically-based analytical  expression is tested both in models with and without buoyancy glitches. It does not assume that glitches are small and, consequently, predicts non-sinusoidal glitch-induced period-spacing variations, as often seen in model and real data.  {We show that the glitch position and width inferred from the fitting of the analytical expression to model data consisting of pure gravity modes are in close agreement (typically better than 7$\%$ relative difference) with the properties measured directly from the stellar models.} In the case of fitting mixed-mode model data, the same expression is shown to reproduce well the numerical results, when the glitch properties are known a priori. In addition, the fits performed to mixed-mode model data reveal a frequency dependence of the coupling coefficient, $q$, for a moderate-luminosity red-giant-branch model star. Finally, we find that fitting the analytical expression to the mixed-mode period spacings may provide a way to infer the frequencies of the pure acoustic dipole modes that would exist if no coupling took place between acoustic and gravity waves.
\end{abstract}




\section{Introduction\label{introduction}}


Stellar oscillations provide a direct probe of the chemical gradients
inside stars caused by different physical
processes such as
nuclear burning, microscopic diffusion, and macroscopic mixing, in,
and beyond, the convectively unstable regions
\citep[e.g.][]{bossini15,constantino15, pedersen18}.  With the advent
of space missions with programmes dedicated to the observation of
stellar oscillations, such as CoRoT~\citep{baglin06} and Kepler
\citep{gillilandetal10}, the opportunity to use ultra-precise
and abundant seismic data to constrain these physical processes has
flourished, establishing new challenges also for the understanding of the
relation between the details of the stellar structure and the
signatures imprinted by these details on the seismic data
\citep[e.g.][for a recent review]{hekker17}. In this
context, the study of internal gravity waves and waves of mixed nature are of
particular relevance.

Internal gravity waves are observed in intermediate to high mass
pulsators, subdwarf B stars, and white dwarfs. In addition, in
subgiant and red-giant stars, waves of mixed nature may be observed, which have
the properties of a gravity wave in the inner radiative layers and the
properties of an acoustic wave in the stellar envelope. 
The internal gravity waves are maintained by
gravity acting on density fluctuations, have frequencies
below the buoyancy frequency, and propagate in non-convective
regions only. Their propagation speed depends directly on the
buoyancy frequency, defined by
\begin{eqnarray}
N^2=g\left(\frac{1}{\gamma_1}\frac{\rmd\ln p}{\rmd r}-\frac{\rmd \ln\rho}{\rmd r}\right),
\label{bruntdef}
\end{eqnarray}
where $g$ is the gravitational acceleration, $\gamma_1$ is the first
adiabatic exponent, $p$ is the pressure, $\rho$ is the density, and
$r$ is the distance from the stellar centre. 

Asymptotically, the oscillation periods of eigenmodes of gravity
nature (hereafter, g modes) are approximately equally
spaced. Consequently, the difference between two modes of the same degree, $l$, and
consecutive radial orders, $n$, known as the period spacing, $\Delta
P$,  is approximately constant. This asymptotic value of the
period spacing  is given by \citep{tassoul80,aerts10},
\begin{equation}
\Delta P_{\rm as}= \frac{2\pi^2}{\omega_{\rm g}},
\label{psasymp}
\end{equation}
where,
\begin{equation}
\omega_{\rm g}\equiv\int_{r_1}^{r_2}\frac{LN}{r}\rmd r,
\label{omegag}
\end{equation}
$L^2=l(l+1)$, and $r_1$ and $r_2$ are the inner and outer turning
points, respectively, that define the propagation cavity of the g mode. 

The above assumes a spherically symmetric stellar equilibrium, thus, it neglects the potential impact of rotation on the oscillations.  This can be critical, particularly when considering intermediate to high-mass pulsators which typically rotate fast \cite[see,][for a recent review]{aertsetal18}. We shall keep this assumption throughout the paper. However, given the importance of rotation for pulsators in particular regions of the HR diagramme, that effect shall be considered in a follow-up work. 

Sharp variations in the buoyancy frequency inside the g-mode
propagation cavity may deflect the oscillation periods from their
asymptotic values. This happens when the scale of variation of $N$ is
comparable to, or smaller than the local wavelength of the wave. This kind
of variations, known as structural (buoyancy) glitches,  cause the period
spacing to deviate from the constant asymptotic
value. These glitches are associated with strong gradients in
chemical composition, resulting from a combination of physical
processes, such as nuclear
burning, diffusion, and mixing, and may be found at different locations, including at
some borders, or former borders, of convective 
regions and in nuclear burning shells. 

The impact of a buoyancy glitch on the period spacing depends strongly
on the position of the glitch in the propagation cavity of the
g mode. That position is best measured in terms of the buoyancy
radius\footnote{We note that in \cite{cunha15} we have mentioned that this definition was different from that in ~\cite{miglio08}. In fact, their definition is entirely consistent with ours, the only difference being that we opted to include $L$ in our definition, while they do not do so.}, defined by
\begin{equation}
\tilde\omega_{\rm g}^{r}=\int_{r_1}^{r}\frac{LN}{r}\rmd r,
\end{equation}
or the buoyancy depth, defined by
$\omega_{\rm g}^r\equiv\omega_{\rm g}-\tilde\omega_{\rm g}^r$.
The closer the glitch is to the middle of the propagation cavity
(defined by $\tilde\omega_{\rm g}^r/\omega_{\rm g} = 0.5$), the shorter is the scale in which the
period spacing varies with frequency. For the remaining of this paper,
we shall refer to the inner half of the gravity wave propagation
cavity as the region
where $\tilde\omega_{\rm g}^r/\omega_{\rm g} < 0.5$ and to the outer half as the
region where  $\tilde\omega_{\rm g}^r/\omega_{\rm g} > 0.5$.

In the case of red-giant stars, where pulsations have a mixed nature,
the characteristic pulsation frequency
spectrum shows signatures of both gravity and acoustic
pulsation spectra. Since the oscillations are driven by convection,
the oscillation power is modulated by an envelope centred around
the frequency of maximum power $\nu_{\rm max}$, that can be scaled from
the solar case \citep{brown91,kjeldsen95}. Moreover, the period
spacing follows approximately the asymptotic expectation for g modes
for frequencies significantly different from what would be the
frequencies of pure acoustic modes in the star  (i.e., eigenmodes of pure acoustic
nature, hereafter, p modes). However, close to the
pure acoustic frequencies,  there is a strong coupling between the oscillation in the inner (g) and outer
(p) cavities and the period spacing decreases significantly with
respect to the asymptotic value. These dips in the period spacing are approximately
equally spaced in frequency, by the  large frequency separation, whose first-order
asymptotic value is given by \citep{tassoul80,gough93}
\begin{equation}
\Delta\nu_{\rm as}=\left(2\int_0^Rc^{-1}{\rmd r}\right)^{-1},
\label{deltanu_as}
\end{equation} 
where $c$ is the sound speed.

The impact of buoyancy glitches on the periods of g modes has been
theoretically addressed in previous works in the context
of the study of white dwarfs and main-sequence intermediate-mass
stars \citep[e.g.][]{brassard92,miglio08,wu18}. However, no explicit expression for the period spacing variation was presented by these authors, except for the case of the small glitch limit, when the variation is sinusoidal. Likewise, the frequencies
of mixed modes in red-giant stars have been modelled by
\cite{mosser12}, based on the asymptotic work by
\cite{shibahashi79} and \cite{unno89}. An explicit expression for the period spacing variations was presented in \cite{mosser15}, but in a form that requires an interpolation procedure. An equivalent formulation was simultaneously presented by \cite{cunha15}, and again, independently, by \cite{hekker17}, that does not require such interpolation. \cite{cunha15} have also studied
the combined effect of buoyancy glitches and mode coupling in mixed modes,
deriving an explicit, asymptotically-based analytical expression for
the period spacings where these effects are accounted for.
However, in that work the authors addressed only the case of a glitch
modelled by a Dirac delta function, which does not reproduce well the
variety of glitch shapes that is found in stellar models (potentially also in real data). Moreover, they concentrated on the case of
red giants, for which all observed modes are of mixed nature. 

In the present work we investigate the impact of structural glitches on the properties of stellar oscillations further, by
considering glitches of different shapes and the signatures they introduce both on mixed
modes and on pure gravity modes. We stress that our approach does not assume that the glitch is small and, as a consequence, does not lead to sinusoidal period-spacing variations, except in that limit. In addition, we re-visit the
asymptotic description for the case when
coupling between acoustic and gravity waves occurs, but no glitch is present in the
g-mode cavity. In particular,  we demonstrate that the analytical
expression proposed by \cite{cunha15}, now extended in the way
discussed in the subsequent sections of this paper, reproduces well the
period spacings computed from model data and that it can be used
to: (1) model the impact of buoyancy glitches on pure g modes in
stars where they
are observed, {\it e.g.}, main-sequence intermediate-mass
stars, subdwarf B stars, and white dwarfs, (2) model the 
coupling between the g and p modes in the absence of glitches in
red-giant stars, and (3) model the combined effect of the glitches and
coupling on red-giant mixed modes. The general analytical
expression for the period spacing is presented in
Sec.~\ref{general}. In Secs~\ref{g}-\ref{g_c} this expression is
tested against model data for three different cases, namely, a case
of a glitch and no coupling (so, pure g modes), a case of coupling and
no glitch (the typical mixed modes), and a case of combined
glitch and coupling effects on mixed modes. In Sec.~\ref{conclusion} we discuss our
results and conclude. The details of the analytical derivations are
provided in Appendix~\ref{apa}.  

\section{General analytical formulation and models}
\label{general}

The starting point for the work presented here is the expression for
the relative period spacing, \dP$/\Delta P_{\rm as}$, in the presence of 
mode coupling and a buoyancy glitch presented by
\cite{cunha15}, according to which,
\begin{equation}
\frac{\Delta P}{\Delta P_{\rm as}}\approx \frac{1}{1-\displaystyle({\omega^2}/{\omega_{\rm
    g}})\left[{\rmd\varphi}/{\rmd\omega}+{\rmd\Phi}/{\rmd\omega}\right]},
\label{ps_coupling_glitch}
\end{equation}
where we recall that the asymptotic period spacing $\Delta P_{\rm as}$ on the left hand side can be expressed in terms of the buoyancy size of the g-mode cavity, $\omega_{\rm g}$, following Eq.~(\ref{psasymp}). The coupling phase, $\varphi$, incorporates the effect of the coupling between p and
g modes on the period spacing and is given by
equation~34 in \cite{cunha15}, namely,
\begin{equation}
\varphi={\rm atan}\left[\frac{q}{\tan\left[\left(\omega-\omega_{{\rm a},n}\right)/\omega_{\rm p}\right]}\right],
\label{varphi}
\end{equation}
where
\begin{equation}
\omega_{\rm p}=\left(\int_{r_3}^{r_4}c^{-1}\rmd r\right)^{-1},
\label{omega_p}
\end{equation} 
$r_3$ and $r_4$ are the turning points of the p-mode cavity, and $q$ is the coupling coefficient~\citep{unno89,takata16}.
Also, $\omega_{{\rm a},n}$ is the angular frequency of what would be the pure
acoustic mode of (pressure) radial order $n$, in the absence of mode
coupling.
The glitch phase, $\Phi$, incorporates
the effect of the structural buoyancy glitch. Both phases are frequency
dependent. Moreover, in the most general case, the glitch phase $\Phi$
depends on the coupling phase
$\varphi$.  This is because the impact of the structural glitch on
the oscillation period depends on the phase of the wave at the glitch
position and that phase, in turn, may depend on the mode coupling. We further note, from inspection of Eqs~(\ref{omega_p}) and (\ref{deltanu_as}), that the quantity $\omega_{\rm p}$ is related to the model asymptotic large frequency separation by
$
\omega_{\rm p}\approx 2\Delta\nu_{\rm as},
$
 with the restriction that the left-hand side is never smaller than the right-hand side term. Moreover, while $\omega_{\rm p}$ may depend on frequency, through a possible frequency dependence of the turning points, $\Delta\nu_{\rm as}$ is fully defined by the equilibrium model, hence, is, by definition, frequency independent. 

Analytical expressions for the coupling and glitch phases were presented in
\cite{cunha15} for the case of a
glitch modelled by a Dirac delta function located in the outer half of
the g-mode cavity.  Here we shall present, in addition, formulations for the
cases of glitches modelled either by a step function or by a
gaussian-like function, which describe more adequately the types of structural
variations that are seen in the stellar models considered in this work. 


In the absence of a structural glitch, $\Phi=0$. Then, 
eq.~(\ref{ps_coupling_glitch}) reduces to,
\begin{equation}
\frac{\Delta P}{\Delta P_{\rm as}}\approx \frac{1}{1-\displaystyle({\omega^2}/{\omega_{\rm
    g}})\left({\rmd\varphi}/{\rmd\omega}\right)} \equiv\zeta(\omega),
\label{ps_coupling}
\end{equation}
an expression that was first presented in~\cite{jcd12},  with the
explicit form of $\varphi$ given later by \cite{cunha15} ({\it
  cf.} Eq.~(\ref{varphi})). In
\cite{mosser15} this relative bumped period spacing was identified with the function $\zeta(\omega)$  defined by
\cite{deheuvels15} (following on the work by \cite{goupil13}) in the context of the study of mixed-mode
rotational splittings. The frequency position of the acoustic resonances,
characterized by an abrupt decrease of the period spacing,  corresponds
to the minima of the function $\zeta$, and, in turn, to the maxima
of $-{\rm d}\varphi/{\rm d}\omega$.  We shall see, in
Sec.~\ref{c}, that the analytical expression for the relative
period spacing presented by \cite{cunha15} for the case of coupling and no glitch is
equivalent to the function $\zeta (\omega)$, but that it is written in such a way that it is much easier to fit to real data than the version
presented by \cite{deheuvels15}. 

It is important to note that the impacts on the period spacing from the
mode coupling and from a
structural glitch are, generally, not additive. When the glitch effect
is small, meaning
\begin{equation}
\left|1-\displaystyle\frac{\omega^2}{\omega_{\rm 
    g}}\frac{\rmd\varphi}{\rmd\omega}\right| \gg \left|\displaystyle\frac{\omega^2}{\omega_{\rm
    g}}\frac{\rmd\Phi}{\rmd\omega}\right|,
\label{small}
\end{equation}
eq.~(\ref{ps_coupling_glitch}) can be approximated by,
\begin{equation}
\frac{\Delta P}{\Delta P_{\rm as}}\approx \zeta\left(\omega\right)+\frac{\omega^2}{\omega_{\rm
    g}}\frac{\rmd\Phi}{\rmd\omega}.
\label{dpovp}
\end{equation}
In this limit case  (eq.~(\ref{dpovp})),  the relative period-spacing variation is found to be similar to that presented by \cite{mosser15}
(their equation 28).
 However, even in this case there is an important
difference between the two results that is worth noting: in eq.~(\ref{dpovp})  the glitch
term (second term on the right hand side (rhs)) generally depends on the
coupling term, through the dependence of the glitch phase
$\Phi$ on the coupling phase $\varphi$, while that fact was
not considered in the
work of \cite{mosser15}. As briefly discussed by \cite{cunha15}, the
fact that $\Phi$ generally depends on $\varphi$ has significant implications for the combined
period-spacing modulation at the acoustic resonances, requiring that
the two effects are
modelled simultaneously, rather than sequentially.

In this work, stellar models will be used to test the 
analytical expression given by eq.~(\ref{ps_coupling_glitch}) in its
various forms described in detail in Secs~\ref{g}-\ref{g_c}. The models, whose global properties are summarised in Table~\ref{model_parameters}, are computed with the evolution code
\astec~\citep{jcd08a} and the corresponding pulsation frequencies are computed with the adiabatic pulsation
code \adipls~\citep{jcd08b}. For the study of the glitch effect on the periods of pure gravity
waves we consider a main-sequence stellar model with a mass
$M=6$\msol\, (Sec.~\ref{g}) and a 1\msol\ red-giant-branch (RGB) model located at the luminosity
bump \citep[the same as Model 1a in][]{cunha15}. For the latter we used the ASTER code\footnote{The
	ASTER code computes the solutions to the adiabatic pulsation
	equations under the Cowling approximation considering only the g
	modes. This is done by taking the local radial wavenumber $K$ to be defined by the relation $K^2=-(L^2/r^2)(1-N^2/\omega^2)$} to compute the frequencies of what would be the
pure g modes if no coupling existed, by artificially disregarding the p-mode cavity, as
explained in \cite{cunha15} (their section 3.2.3). The effect of
the mode coupling in the absence of structural glitches is tested on
a 1\msol\ RGB stellar model with a luminosity that
is lower than the luminosity bump  (Sec.~\ref{c}). Finally, the
combined effect of mode coupling and a buoyancy glitch is tested on the
1\msol\ RGB model located at the luminosity bump (Sec.~\ref{g_c}),
using the mixed-mode frequencies computed with ADIPLS. The choice of this latter RGB model is motivated by the fact that a clear buoyancy glitch is found at that luminosity. Thus, even if mixed modes may be harder to detect observationally in such high-luminosity RGBs \citep{mosser18}, the analytical expression is best validated in such a clear case.

\begin{table*}
	\caption{Properties of the stellar models considered in this work. The frequency of maximum power for the lower luminosity and higher luminosity RGB models are, respectively, $\nu_{\rm max}$=105~$\mu$Hz and $\nu_{\rm max}$=40~$\mu$Hz 
	}
		\begin{minipage}{0.7\textwidth}
			\resizebox{\linewidth}{!}{%
				\begin{tabular}{|l|c|c|c|c|c|}
					\hline
					{\bf Model} & {\bf Mass} (\msol) & {\bf Radius} (\rsol) & {\bf Effective Temp.} (K) & {\bf Luminosity} (\lsol) & {\bf Age} (Gyr) \\
					\hline
					Main sequence &6.0 & 3.5 & 18661& 1328 &0.0236\\
					\hline
					RGB-1 (no core glitch) &1.0& 5.8 & 4624 & 14.0  & 11.37 \\
					\hline
					RGB-2 (core glitch)& 1.0 & 9.7 & 4438 & 32.7 &  11.45\\
					\hline
				\end{tabular}
			}
		\end{minipage}
		\label{model_parameters}
	\end{table*}

\section{Buoyancy glitch effect on pure gravity waves}
 \label{g}
The impact of buoyancy glitches on pure gravity waves has been
addressed through asymptotic analysis in previous theoretical works related to white dwarfs \citep{brassard92} and relatively massive main-sequence
stars \citep{miglio08}.
 In both cases the glitch was assumed to be
well described by a step function and no explicit expression for the
period-spacing modulation was presented, except in the limit case of small
glitches.
In
\cite{cunha15} an explicit expression for the period-spacing
modulation was provided for the effect of a glitch on pure gravity
waves, but only for the case of a glitch modelled by a Dirac
delta function \footnote{We note that in that paper there is a typo in
  equation (25): a minus sign should have preceded the
  expression for $\mathcal{F}_G$. The same sign is missing in the second
  term on the right hand side of equation (39) of the same paper. Nevertheless, all results
  presented in that work have considered the correct sign and
  are, therefore, correct.}. Since that work concerned only red giant stars, the
analytical expression was tested on models by computing the pulsation
equations with modified boundary conditions that assumed that no p modes were present, to avoid
the effect of mode coupling. It was found that, except for the
amplitude dependence on frequency, the analytical expression provided
a good fit to the numerical results. The mismatch between the
frequency dependences of the analytical and numerical amplitudes found
by the authors is explained by the fact that in
the numerical model the glitch presented a finite width, which was not
accounted for when modelling it with a Dirac delta function.
To overcome that limitation, here we model again the glitch seen in that RGB
model \citep[Model 1a in][]{cunha15} but using, instead, a Gaussian-like function. Moreover, for the case of the
main-sequence intermediate-mass model, the glitch will be modelled with a step-like function,
as detailed below.

\begin{figure*} 
         \begin{minipage}{1.\linewidth}  
               \rotatebox{0}{\includegraphics[width=0.49\linewidth]{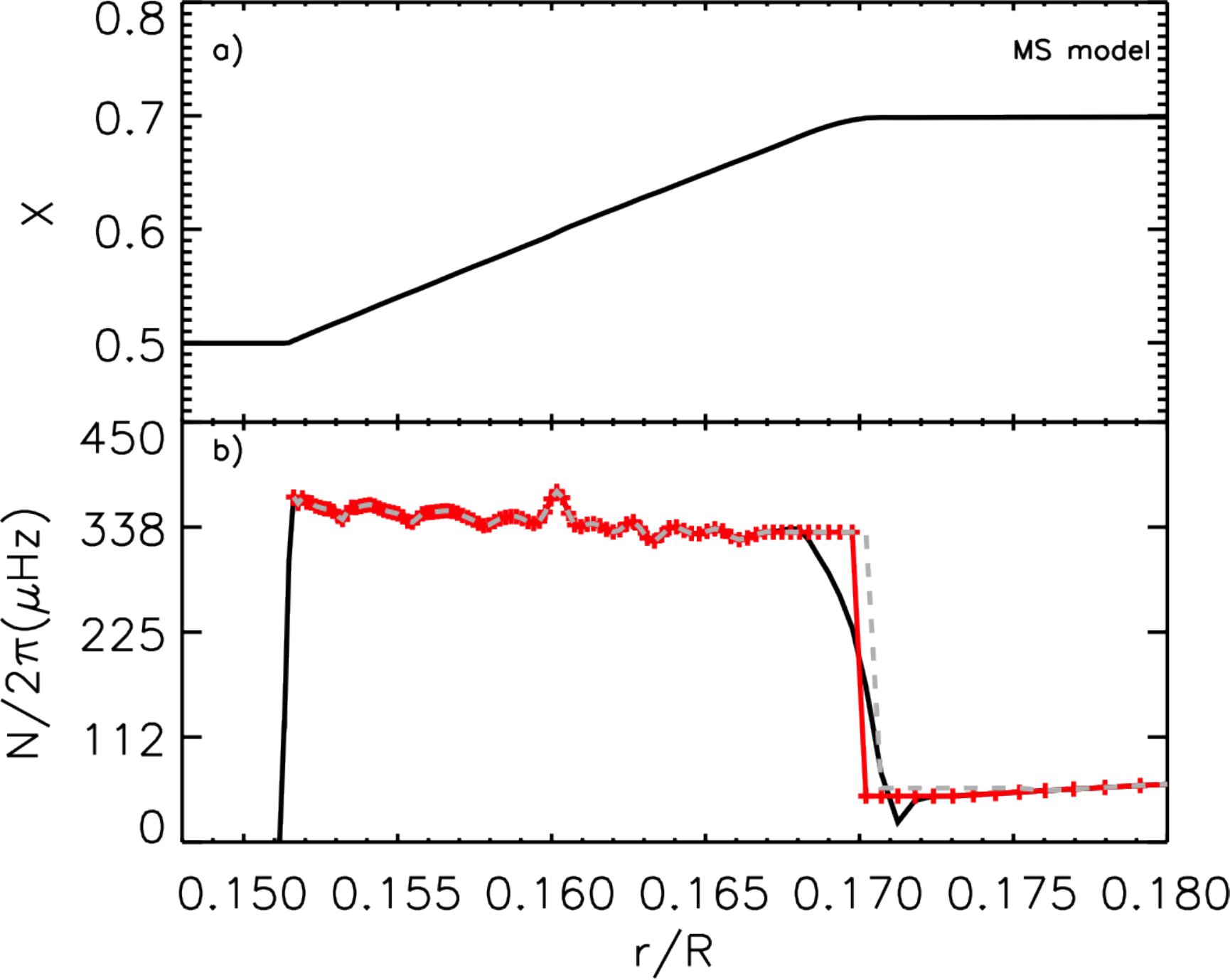}}
              \rotatebox{0}{\includegraphics[width=0.49\linewidth]{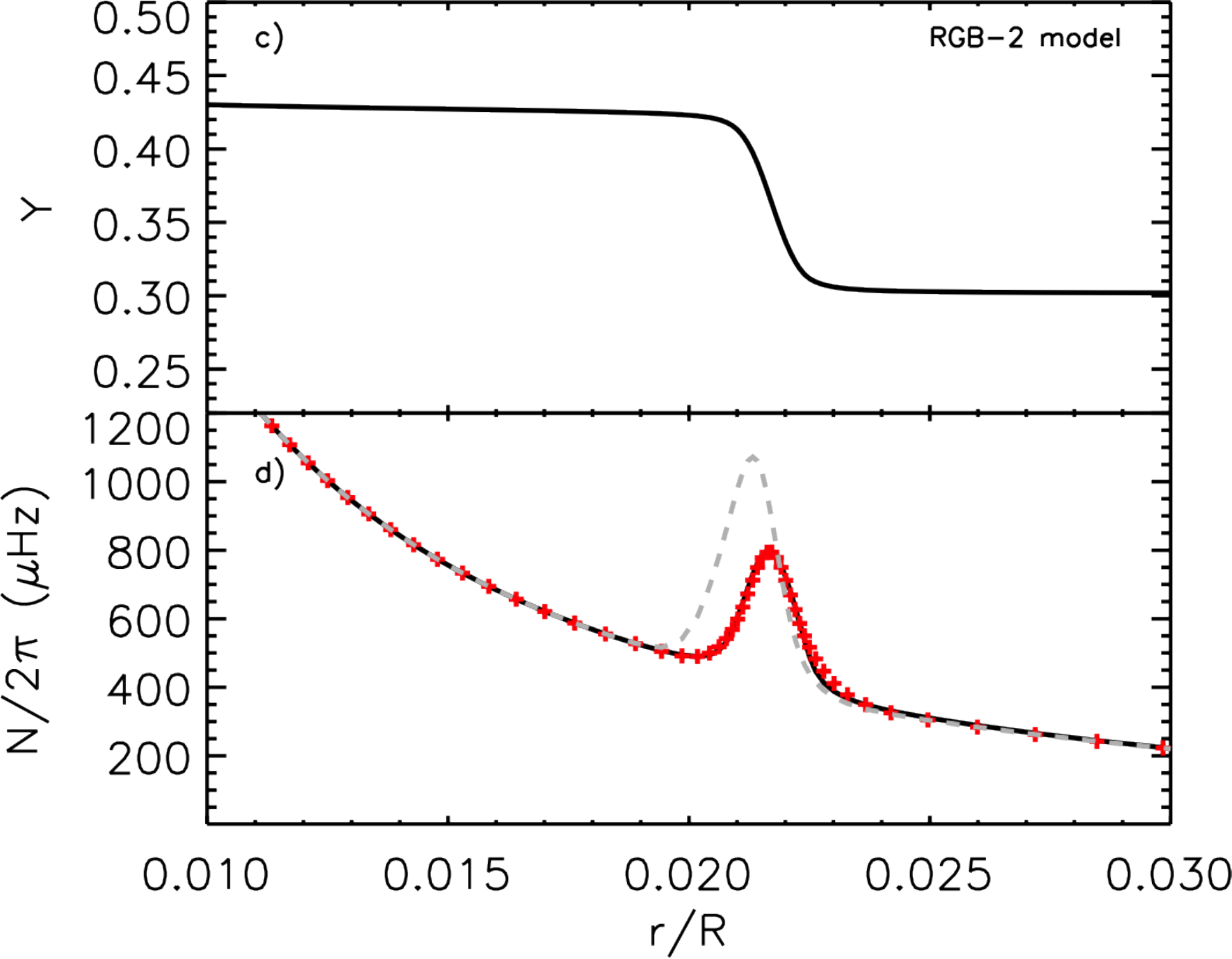}}
         \end{minipage}
       \caption{Hydrogen profile (panel a)) and buoyancy frequency (panel b)) for the 6M$_\odot$ main-sequence model and helium profile (panel c)) and buoyancy frequency (panel d))  for the RGB model. The figures show the regions where the
         glitches are located in these models. The black curves show the results from ASTEC while the red
       crosses show the models described by
       Eqs~(\ref{brunt}) (panel b)) and (\ref{glitch}) (panel d)) used to
       derive the analytical expressions for the relative 
       period-spacing variation. {The grey, dashed lines in the two bottom panels show the buoyancy frequency recovered from the same models when adopting the parameters inferred from the fit of the analytical expressions to the numerical period spacings.}
}
      \label{fig:glitch}
       \end{figure*}

The bottom panels of Fig.~\ref{fig:glitch} show the buoyancy
frequency  around
the position of the glitch for the two models considered in this
section, as a function of the relative distance from the centre
of the star, $r/R$, where $R$ is the stellar radius. In the 6~\msol, main-sequence stellar
model, the buoyancy frequency  (Fig.~\ref{fig:glitch}\,b) drops abruptly at $r^\star/R=0.17$, from the plateau resulting from the steep slope in the hydrogen profile outside the retracting convective core (Fig.~\ref{fig:glitch}\,a).  We note that no smoothing of the composition profile was considered during the evolution of this model. The drop in the buoyancy frequency  can be modelled by the discontinuous function,
\begin{equation}
N=\left\{
\begin{array}{lll}
 N_{\rm in}  &   {\rm for} & r < r^\star\\
 N_{\rm out}  & {\rm for} & r > r^\star \\
\end{array}
\right. ,
\label{brunt}
\end{equation}
with $N$ varying by $\Delta N= \left. N_{\rm in}\right|_{ r\rightarrow r^\star_-}-\left. N_{\rm out}\right|_ {r\rightarrow r^\star_+}$ at $r=r^\star$.
The glitch is thus characterised by two parameters,
namely, the relative step amplitude
$A_{\rm st}=[N_{\rm in}/N_{\rm out}]_{r^\star}-1$, and the position, $r^\star$.

In the case of the 1~\msol\ red-giant model, the buoyancy frequency (Fig.~\ref{fig:glitch}\,d) shows a hump at  $r^\star/R=0.0216$. This hump results from the
strong chemical gradient generated at the time of the first dredge-up and left behind by the retreating
convective envelope (Fig.~\ref{fig:glitch}\,c). Here, numerical diffusion associated to the treatment of the
	mesh in the stellar evolution code leads to a smoother composition profile, hence also to a broader feature in the  buoyancy frequency.  We
model this glitch using a Gaussian-like function by defining,
{
\begin{eqnarray}
{N}={N_0}\left[1+\frac{A_{\rm G}}{\sqrt{2\pi}\Delta_{\rm g}}\exp{\left(-\frac{(\omega_{\rm g}^r-\omega_{\rm g}^{\star})^2}{2\Delta_{\rm g}^2}\right)}\right],
\label{glitch}
\end{eqnarray}
where  $N_{0}$ is the glitch-free buoyancy frequency. In this case,
the glitch is characterised by three parameters, the constants $A_{\rm G}$
and $\Delta_{\rm g}$, which measure, respectively, the amplitude and width of
the glitch, and the glitch position $r^\star$, which enters the buoyancy depth at the glitch position,
$\omega_{\rm g}^{\star}=\int_{r^\star}^{r_2}(LN/r){\rm d}r$. }
This is in contrast to the model assumed in~\cite{cunha15}, where
the glitch was 
 characterised by two parameters only, namely, the amplitude
$A_\delta$ and the position $r^\star$. 


\begin{table*}
	\caption{{Parameters derived from the fit of the analytical expression
			for the step-like glitch (Eq.~(\ref{ps_glitch_st_in}))  to the 
			period spacing derived from ADIPLS for the main-sequence model. Their distributions are shown in Fig.~\ref{caso1_st}. The values shown correspond to the median of the distributions and the $68\%$ confidence intervals. For
			comparison, the values of the glitch parameters estimated directly
			from the buoyancy frequency obtained with ASTEC
			are also shown. The glitches reconstructed from the inferred and estimated parameters are compared in Fig.~\ref{fig:glitch}\,b.}}
	\label{parameters_step}
\label{parameters_step}
\begin{minipage}{0.7\textwidth}
	\resizebox{\linewidth}{!}{%
		\begin{tabular}{|c|c|c|c|c|}
			\hline
			&$\Delta P_{\rm as}$ (s) & $A_{\rm st}$  &$\tilde\omega_{\rm g}^\star$ ($10^{-6}$ rad/s) & $\delta$\\
			\hline
			& & & & \\
			Fit&{$8472^{+50}_{-50}$}&{$4.74^{+0.44}_{-0.39}$} &{$351.61^{+0.73}_{-0.72}$}  &{$0.602^{+0.019}_{-0.019}$} \\
			& & & & \\
			\hline
			Estimated &--&{$5.3$}  & {349}&--\\
			\hline
		\end{tabular}
	}
\end{minipage}
\end{table*}

\begin{figure*} 
	\begin{minipage}{1.\linewidth}  
		\rotatebox{0}{\includegraphics[width=1.0\linewidth]{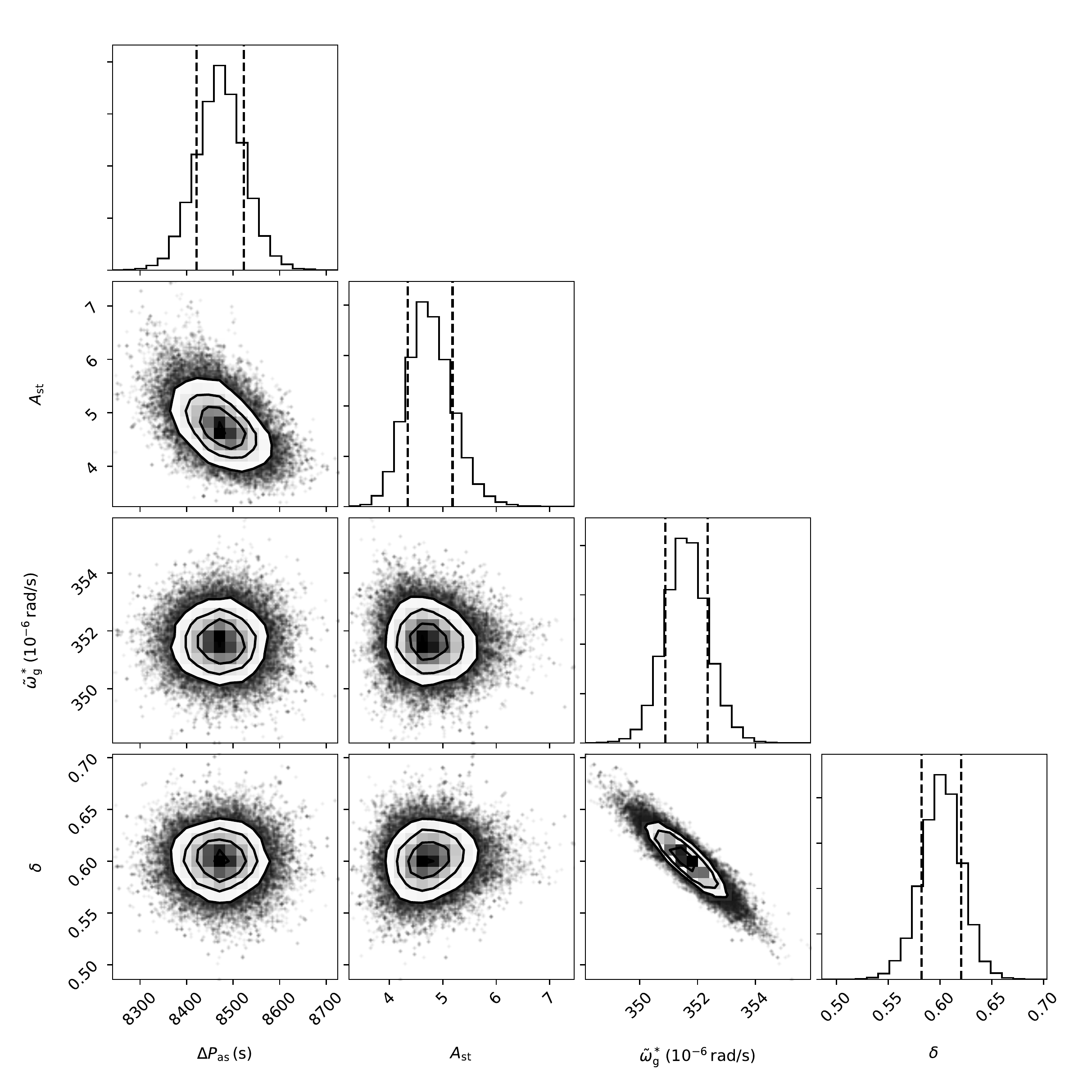}}
	\end{minipage}
	\caption{Marginalised distributions for the parameters considered in
		the fit of the rhs of Eq.~(\ref{ps_glitch_st_in}) to the period
		spacing derived from ADIPLS for the main-sequence model.}
	\label{caso1_st}
\end{figure*}

{The derivation of the eigenvalue condition in the presence of a glitch that leads to the definition of the phase $\Phi$  for each case described above} is carried out in a way similar to that
presented in \cite{cunha15} for the case of a glitch modelled by a
Dirac delta function. The details are presented in Appendix~\ref{apa}. For each glitch considered, we differentiate the glitch
phase (given by Eq.~(\ref{Phi_st_in}) for the step-like glitch and by Eq.~(\ref{Phi_g2}) for the Gaussian-like glitch), introduce it into
Eq.~(\ref{ps_coupling_glitch}), and take ${\rm d}\varphi/{\rm d}r=0$ (in accordance with the no mode coupling
assumption made in this section), to obtain the corresponding period spacing.

{For a glitch modelled by a step function and located in the inner half of
the cavity (as in the main-sequence model considered here) it follows that }\footnote{The signature on the period spacing from a step-like glitch depends on the side of the cavity where it is located. Deriving the expression for the case of a glitch located in the outer half of the cavity is straightforward following the same steps as in Appendix~\ref{apa}.} 
\begin{eqnarray}
\frac{\Delta P}{\Delta P_{\rm as}}\approx \left[1-{\scriptsize{\frac{\tilde\omega_{\rm g}^\star}{\omega_{\rm g}}\;\frac{-A_{\rm st}\sin\tilde\beta_1+A_{\rm st}^2\cos^2\tilde\beta_2}{(1+A_{\rm st}\cos^2\tilde\beta_2)^2+(0.5A_{\rm st}\cos\tilde\beta_1)^2}}}\right]^{-1}\hspace{-0.4cm},
\label{ps_glitch_st_in}
\end{eqnarray}
where $\tilde\beta_1=2\tilde\omega_{\rm g}^\star/\omega+2\delta$ and $\tilde\beta_2=\tilde\omega_{\rm g}^\star/\omega+\pi/4+\delta$. 
{Here, quantities marked with a superscript $^\star$ refer to values
taken at $r=r^\star$ and $\delta$ is a phase related to the details of the mode
reflection near the turning points of the propagation cavity (see
Appendix~\ref{apa} for details).}

 {For the glitch modelled by a Gaussian-like function the derivation of the eigenvalue condition is not as straightforward as for the cases of glitches modelled by a Dirac delta or a step function. The reason is that for the Gaussian-like glitch the derivation requires knowledge of the eigenfunction inside the glitch. As the asymptotic approximation breaks down when the background varies on scales comparable with or smaller than the local wavelength, the asymptotic solution is unlikely to provide an adequate description of the eigenfunction inside the glitch.  For a small enough glitch, this problem can be overcome by making use of the variational properties of the solutions that allow us to derive the perturbation to the oscillation periods without explicitly taking into account the perturbation to the eigenfunctions. However, for a glitch such as the one considered here, that option is not available and proceeding with the derivation of the eigenvalue condition and glitch phase requires a somewhat arbitrary choice for the description of the eigenfunction inside the glitch. We have considered two different options for that choice that we discuss in detail in Appendix~\ref{apa}.} 
 	{We have tested both cases against the limit of a small glitch modelled by a Gaussian-like function, which can be derived without explicit knowledge of the wave solution. Both cases reproduce the functional form derived in the small-glitch limit, but with a frequency attenuation of the glitch signature that differs from that found in the limit case. In the limit of a small glitch, the perturbation to the periods derived from the variational principle varies exponentially as ${\exp}{({-2\Delta_{\rm g}^2\omega^{-2}})}$, whereas the derivation made in Appendix~\ref{apa}, which takes the eigenfunction inside the glitch explicitly into account, predicts that the perturbation to the periods varies as ${\exp}{({-0.5\Delta_{\rm g}^2\omega^{-2}})}$.  Clearly, the expression for a glitch of arbitrary amplitude must reproduce the expression valid in the limit of a small glitch, so it is reasonable to conclude that the difference found results from the inadequate modelling of the eigenfunction inside the glitch. Acknowledging  that the very nature of the asymptotic analysis used in our derivation precludes us from improving it, we have changed the factor in the exponential function to insure  that the analytical expression representing the perturbation induced by a glitch of arbitrary strength satisfies the result found in the small-glitch limit, and then tested the modified analytical expression against numerical results. The results from these tests, detailed in Appendix~\ref{apa}, support the change in the factor introduced in the exponential function, indicating that such change is necessary also when the glitch is strong. With the correction mentioned above, both analytical expressions derived in the Appendix~\ref{apa} provide an adequate fit to the numerical results, although significantly different  amplitudes are recovered from the two fits. Here we discuss the analytical expression that was found to perform best against the numerical results, leaving the detailed comparison with the other case to Appendix~\ref{apa}.}
 	
{Considering a glitch  located in the
outer half of the cavity (as in the case of the RGB model considered here) it follows that the period spacing for the Gaussian-like glitch is given by}\footnote{For a glitch located on the inner half of the cavity the expression would be the same, but with $\omega_{\rm g}^\star$ replaced by $\tilde\omega_{\rm g}^\star$.}  
\begin{eqnarray}
\label{ps_glitch_g}
\frac{\Delta P}{\Delta P_{\rm as}}\hspace{-0.08cm}\approx & &\\
&&\hspace{-1.7cm}\left[1+{\scriptsize{A_{\rm G}
		f_{\omega}^{\Delta_{\rm g}}\frac{\omega_{\rm g}^\star}{\omega_{\rm g}}\frac{
    \left[\cos\beta_1+\left(\omega/\omega_{\rm g}^\star(1-4\Delta_{\rm g}^2/\omega^2)-
      A_{\rm G}
        f_{\omega}^{\Delta_{\rm g}}\right)\sin^2\beta_2\right]}
    {(1-0.5 A_{\rm G}
      f_{\omega}^{\Delta_{\rm g}}\cos\beta_1)^2+(A_{\rm G} f_{\omega}^{\Delta_{\rm g}}\sin^2\beta_2)^2}}}\right]^{-1}\hspace{-0.4cm},\hspace{-0.3cm}\nonumber 
\end{eqnarray}
where we introduced the
frequency-dependent function {$f_{\omega}^{\Delta_{\rm g}}=\omega^{-1}{\rm  e}^{{-2\Delta_{\rm g}^2\omega^{-2}}}$.} Moreover,  here $\beta_1=2\omega_{\rm g}^\star/\omega+2\delta$ and $\beta_2=\omega_{\rm g}^\star/\omega+\pi/4+\delta$. 


The analytical expressions for the relative period spacing presented
above will be useful for fitting real
data and extracting information about the structural variations. Here we test their
suitability based on fits to model data in the following. In this context
it is important to emphasise that in addition to the glitch parameters
discussed before
(two in the case of the step model and three in the case of the Gaussian  model) these expressions contain also one
global seismic parameter, namely, $\Delta P_{\rm as}= 2\pi^2/\omega_{\rm g}$, and the phase parameter,
$\delta$. Tables~\ref{parameters_step} and \ref{parameters_gau} summarise the values inferred for
the parameters from the fitting of the analytical
expressions to the model data,  for the two glitches considered.  The inferred glitch parameters are to be compared with their
estimated values obtained directly from the buoyancy profiles
(red crosses in Fig.~\ref{fig:glitch}\,b,d).

The rhs of equation  Eq.~(\ref{ps_glitch_st_in})  was fitted to the period spacings computed from the eigenfrequencies
obtained with the pulsation code ADIPLS for the main sequence model, $\Delta P_{\rm ADIPLS}$, 
using the python module {\it emcee} implementation of the affine-invariant ensemble sampler for Markov chain Monte Carlo \citep{emcee} with the 
likelihood defined by
\begin{equation}
{\mathcal L} = \frac{1}{\sqrt{2\pi}\sigma}\exp\left(-\frac12 \chi^2\right),
\label{none}
\end{equation}
where the uncertainty $\sigma$ was left as a free parameter and
\begin{equation}
\chi^2=\sum_i\left(\frac{\Delta P_i-\Delta P_{{\rm ADIPLS},i}}{\sigma}\right)^2.
\end{equation}

\begin{figure} 
         \begin{minipage}{1.\linewidth}  
               \rotatebox{0}{\includegraphics[width=0.97\linewidth]{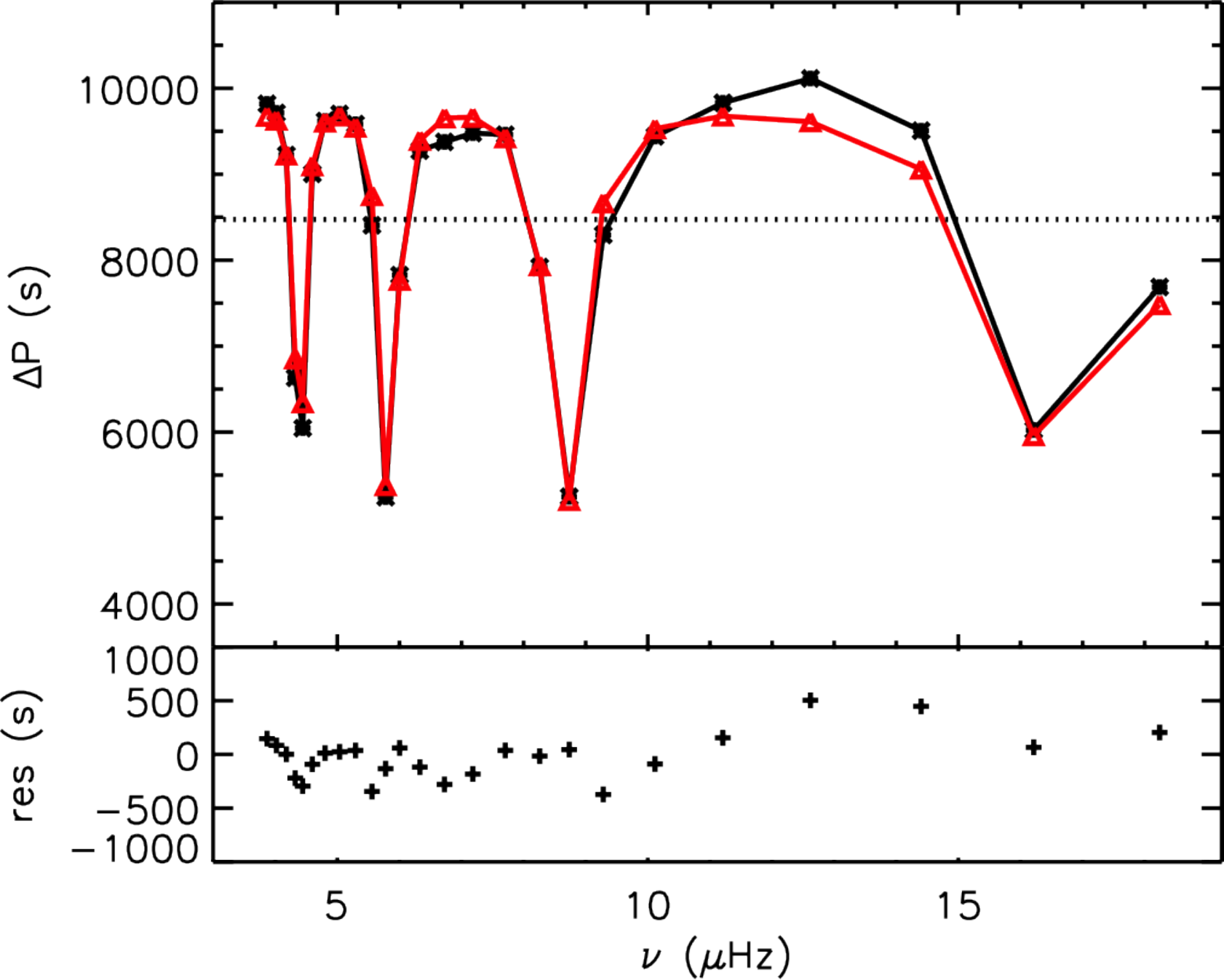}}
         \end{minipage}
\caption{Top panel: comparison between the period spacing derived from ADIPLS
  (black  line and asterisks) for
the main sequence model and that obtained from
Eq.~(\ref{ps_glitch_st_in})  with the most likely parameters from our fit,  performed in the frequency range shown in the figure (red line and triangles). Bottom panel: the residuals (`ADIPLS period spacing'$-$`analytical period spacing')}
   \label{caso1_best_st}
 \end{figure}

 The probability density functions obtained for the parameters
in the fit are shown in Fig.~\ref{caso1_st}. A comparison
of the glitch parameters derived in this way with the values inferred
directly from the buoyancy frequency (Table~\ref{parameters_step}) shows a
reasonable agreement. {While the small differences appear significant, given the errors, they are fully justified by the fact that the step function does not provide an accurate description of the glitch, as seen from Fig.~\ref{fig:glitch}\,b.  In this figure we show, for comparison, the glitch model used to estimate the parameters provided in Table~\ref{parameters_step} (red crosses) and the glitch recovered from the parameters inferred from the fit to the period spacings (dashed, grey line).  The estimated position of the glitch shown in red was taken to be the mid point between the plateaus on the right and left sides of the buoyancy jump. However, given that the jump has a finite extent, the uncertainty associated to this position is more significant than the difference between the estimated and inferred values. Similarly,  Fig.~\ref{fig:glitch}\,b indicates that the difference between the estimated and inferred amplitudes can be accounted for by the deviation of the glitch from a true step function. 
Figure~\ref{caso1_best_st} shows a comparison of the period spacing
computed from the ADIPLS results (black) and that obtained from
Eq.~(\ref{ps_glitch_st_in}) (red) with the parameters of the most likely
model for the fit considered in Fig.~\ref{caso1_st}.}


\begin{table*}
	\caption{{Parameters derived from the fit of the analytical expression
			for the Gaussian-like glitch (Eq.~(\ref{ps_glitch_g}))  to the 
			period spacing derived from ASTER for the RGB-2 model (at the
			luminosity bump). Their distributions are shown in Fig.~\ref{caso1_g}. The values shown correspond to the median of the distributions and the $68\%$ confidence intervals. For
			comparison, the values of the glitch parameters estimated directly
			from the buoyancy frequency obtained with ASTEC
			are also shown. The glitches reconstructed from the inferred and estimated parameters are compared in Fig.~\ref{fig:glitch}\,d.}}
	\label{parameters_gau}
	\begin{minipage}{0.99\textwidth}
		\resizebox{\linewidth}{!}{%
			\begin{tabular}{|c|c|c|c|c|c|}
				\hline
				&$\Delta P_{\rm as}$ (s) & $A_{\rm G}$ ($10^{-6}$ rad/s)  &
				$\omega_{\rm g}^\star$ ($10^{-6}$ rad/s)  & $\Delta_{\rm g}$  ($10^{-6}$ rad/s) &
				$\delta$\\
				\hline
				& & & & &  \\
				{Fit}& {$67.534^{+0.005}_{-0.005}$} & {$607^{+27}_{-25}$} & 
				{$1747.3^{+7.6}_{-7.7}$} &{$158.5^{+3.4}_{-3.4} $}&  {$-0.872^{+0.034}_{-0.034}$}\\
				& & & & &\\
				\hline
				{Estimated}&--&{$380$} & {$1632$} & {$156$} &--\\
				\hline
			\end{tabular}
		}
	\end{minipage}
\end{table*}

    \begin{figure*} 
         \begin{minipage}{1.\linewidth}  
               \rotatebox{0}{\includegraphics[width=1.0\linewidth]{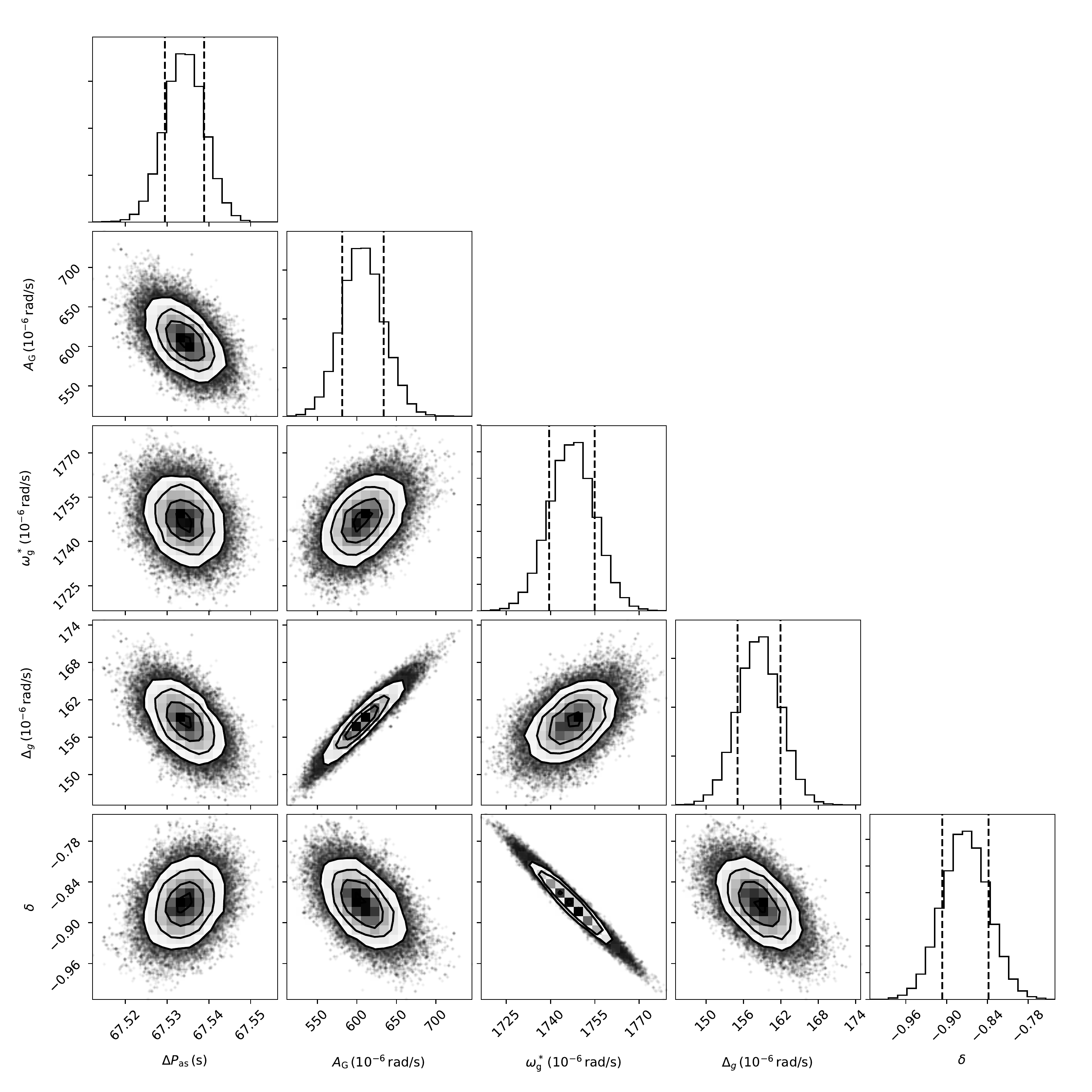}}
         \end{minipage}
\caption{Marginalised distributions for the parameters considered in
  the fit of the rhs of Eq.~(\ref{ps_glitch_g}) to the period
  spacing derived from ASTER for our RGB-2 model (with a core glitch), when coupling
  between the p and g modes is ignored.}
   \label{caso1_g}
    \end{figure*}

    \begin{figure} 
         \begin{minipage}{0.98\linewidth}  
               \rotatebox{0}{\includegraphics[width=1.0\linewidth]{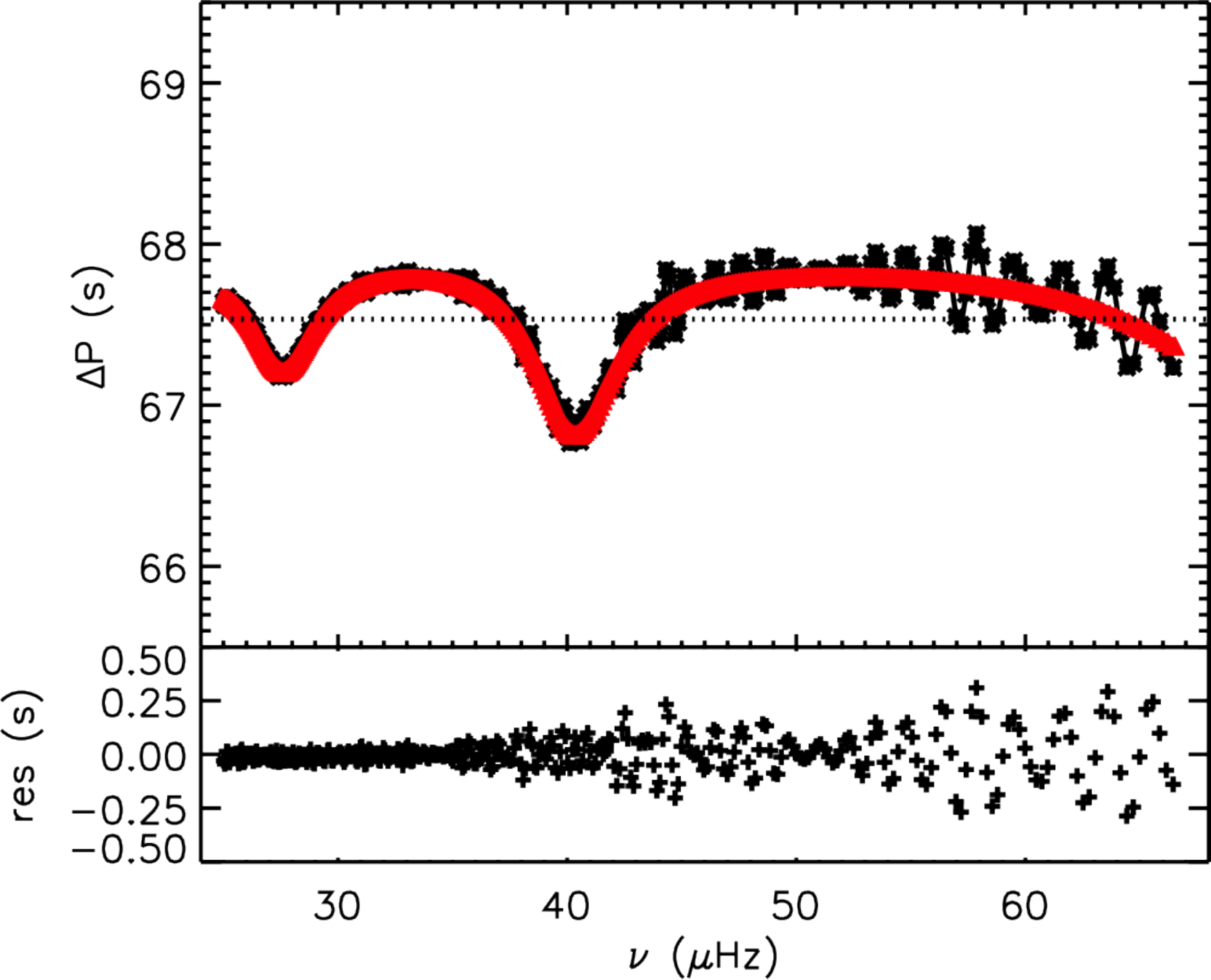}}
         \end{minipage}
\caption{Top panel: comparison between the period spacing derived from ASTER for
our  RGB-2 model with a core glitch (black line and asterisks),  when coupling
  between the p and g modes is ignored, and that obtained from
Eq.~(\ref{ps_glitch_g})  with the most likely parameters from our fit,  performed in the frequency range shown in the figure (red line and triangles). The short scale variations in the black curve result from rapid variations in the second derivative of the buoyancy frequency at the H-burning shell  which are unphysical and, thus, not accounted for in the analytical model.  Bottom panel: the residuals (`ASTER period spacing' $-$ `analytical period spacing').}
   \label{caso1_best_g}
  \end{figure}

To test the analytical expression for the relative period-spacing
variation caused by a Gaussian-like glitch, the rhs of equation
Eq.~(\ref{ps_glitch_g})  was fitted to the period spacings computed
from the eigenfrequencies of pure g modes
obtained with the ASTER code for the RGB-2 model (cf.~ Table~\ref{model_parameters}). The probability density functions obtained for the parameters
entering the fit are shown in Fig.~\ref{caso1_g}. {Figure~\ref{caso1_best_g} shows a comparison of the period spacing
computed from the ASTER results (black) and that obtained from
Eq.~(\ref{ps_glitch_g}) (red) with the parameters of the most likely
model from that fit. }

{A comparison
of the glitch parameters inferred in this way with the values derived
directly from the buoyancy frequency (Table~\ref{parameters_gau}) shows 
that the inferred width of the glitch is in agreement with the estimated one. As for the buoyancy depth  of the glitch, $\omega_{\rm g}^\star$, the value inferred from the fit differs from that estimated by $\sim 7\%$. This difference is significant, given the small  errors, and we have checked that it cannot be explained by an uncertainty in the estimated parameter. In fact, from inspection of Fig.~\ref{fig:glitch}d it seems unlikely that it is related to the modelling of the glitch, which is well represented by Eq.~(\ref{glitch}). On the other hand, we note that the difference is smaller than the width of the glitch.  Given the
approximations made in the course of the derivation of the analytical
expression, the width of the glitch may, in fact, set a limit to  the accuracy with which $\omega_{\rm g}^\star$ can be derived through this method. Finally, the amplitude inferred from the fitting is clearly overestimated, indicating that the predictive power of the analytical expression is more limited for this parameter. The buoyancy glitch that results from assuming the parameters inferred from the fitting is shown in grey on Fig.~\ref{fig:glitch}\,d, for comparison with the glitch seen on the ASTEC model.}

An important aspect to note when comparing the signatures on the
period spacing of glitches modelled by different
functions is that the frequency
dependence of the glitch signature's amplitude ({\it i.e.} the maximum to minimum period-spacing variation induced by the glitch) is different. 
For  the step-like
glitch, we see from Eq.~(\ref{ps_glitch_st_in}) that the amplitude of
the glitch signature on the period spacing is determined by the glitch amplitude, $A_{\rm st}$,
and by the glitch location (implicit in $\tilde\omega_{\rm g}^\star$) both of which are
frequency independent. Thus the signature's amplitude is also
independent of frequency. On the other hand, for the
Gaussian-like glitch, we see from Eq.~(\ref{ps_glitch_g}) that the
amplitude of the glitch signature depends, in addition, on the function
$f_{\omega}^{\Delta_{\rm g}}$. As a consequence, in this case the amplitude
of the glitch
signature decreases with decreasing frequency at low frequencies and with increasing frequency at high frequencies. This difference is not so evident in
Figs~\ref{caso1_best_st} and \ref{caso1_best_g} because the
frequency range shown is relatively small, but it is clear when one
compares the case of the Gaussian-like glitch and the case of the Dirac
delta glitch adopted by \cite{cunha15} (see their Fig. 4).  In the
latter case the amplitude of the glitch signature shows a strong increase with
decreasing frequency. That is the reason why the expression for the glitch modelled by a Dirac delta
adopted in their work did
not reproduce well the signature of the glitch seen in the RGB model.
 In fact, when the oscillation frequency decreases, the characteristic scale of the gravity wave
at the glitch position decreases, and the width of the
glitch eventually becomes comparable with the local wavelength. As a consequence, the
glitch impact on the wave propagation decreases, leading to a
decrease of the amplitude of the glitch effect on the period spacing,
as seen in the Gaussian-like case. However, when the
glitch is modelled by a Dirac delta function,
its width is infinitely small and, therefore, always
infinitely smaller than the local wavelength, preventing the above
effect from taking place. 

Similar differences in the frequency
dependence of the glitch signature's amplitude, according to the
glitch model adopted, have been found also in a number of previous works related to both buoyancy and
acoustic glitches \cite[e.g.][]{monteiroetal94,Houdek07,miglio08}.



\section{Coupling between p and g waves in the absence of glitches}
 \label{c}

In red giants, when no structural glitches are present, the period spacing deviates
 from the asymptotic value due to the coupling
between acoustic and gravity waves in a manner described by Eq.~(\ref{ps_coupling}).
According to \cite{cunha15} in this
case we have, 
\begin{eqnarray}
\label{ps_coupling_2}
{{\frac{\Delta P}{\Delta P_{\rm as}}\Resize{7.5cm}{\approx
      \hspace{-0.08cm}\left[1+{
          {\frac{\omega^2}{\omega_{\rm g}}\frac{q}{\omega_{\rm p}}\left[\sin^2\left(\frac{\omega-\omega_{{\rm a},n}}{\omega_{\rm p}}\right)+q^2\cos^2\left(\frac{\omega-\omega_{{\rm a},n}}{\omega_{\rm p}}\right)\right]^{-1}}}\right]^{-1}}}}
\\
\equiv \zeta(\omega) \hspace{6.7cm}\nonumber
\hspace{-0.2cm},\hspace{-0.1cm}
\end{eqnarray}
where the coupling
coefficient, $q$, is considered to
be independent of the frequency (a condition that will be re-visited below).

We used the analytical expression provided by
Eq.~(\ref{ps_coupling_2}) to fit model data following the same
approach as in Sec.~\ref{g}. In
\cite{cunha15}, a series of 1~M$_{\odot}$ stellar models obtained from evolution
tracks covering the RGB evolution phase was tested for the
presence of structural glitches in the core. Signatures of these glitches were
found only in  models at the luminosity
bump. Thus, a model with luminosity below the bump, extracted from that
series, was chosen to test Eq.~(\ref{ps_coupling_2}) (model RGB-1 in Table~\ref{model_parameters}).

In this case, the parameters to be fitted are the global quantities 
$\Delta P_{\rm as}=2\pi^2/\omega_{\rm g}$ and $\omega_{\rm p}\approx 2\Delta\nu_{\rm as}$,
 the coupling coefficient, $q$, and the pure acoustic frequencies
$\omega_{{\rm a},n}$. However, the asymptotic analysis of the pulsation
equations allows us to estimate $\omega_{{\rm a},n}$ from the frequencies of
radial modes. 
With this in mind, when fitting
Eq.~(\ref{ps_coupling_2}) to model data we have considered two different
options to obtain $\omega_{{\rm a},n}$, both based on the asymptotic
expression for the eigenfrequencies (see Appendix~\ref{apb}, for details), namely: 
\begin{itemize}
\item[1.]  
The frequency $\omega_{{\rm a},n}$ is estimated from the
  frequency of the radial mode
  with the same radial order, $\omega_{{\rm a},n}^0$ through,
\begin{equation}
\omega_{{\rm a},n}=
\omega_{{\rm a},n}^0+\pi\Delta\nu_0+\frac{4\pi^2C}{(\omega_{{\rm a},n}^0+\pi\Delta\nu_0)}, 
\label{omega_a1}
\end{equation}
\item[2.] The frequency $\omega_{{\rm a},n}$ is estimated through the
  relation,
 \begin{equation}
\omega_{{\rm a},n}=2\pi
(n+0.5)\Delta\nu_0+2\pi G_{{\rm a},n}+\frac{4\pi^2C}{(\omega_{{\rm a},n}^0+\pi\Delta\nu_0)}.
\label{omega_a2}
\end{equation}
In both cases, $C$ is a constant
  parameter to be fitted and $\Delta\nu_0$ is the average large frequency separation for
  radial modes in the range of radial orders considered. Moreover, the term
  $2\pi G_{a,n}$ is obtained by linearly interpolating $\omega_{{\rm a},n}^0-2\pi n\Delta\nu_0$,
at  the frequency $\omega_{{\rm a},n}^0+\pi\Delta\nu_0$.
By using either of the options above, we require only a single parameter, $C$, and knowledge of the
radial mode frequencies, to express all $\omega_{{\rm a},n}$. This reduces the total number of parameters in the fit, and, at the same time,
guarantees that the $\omega_{{\rm a},n}$ values are related in a way
that makes physical sense. 
\end{itemize}
The potential advantage of eq.~(\ref{omega_a2}), over  eq.~(\ref{omega_a1}),
is that it accounts for additional frequency
dependences of the eigenfrequencies that are common to modes of degree
$l=0$ and $l=1$, including a possible large-scale frequency variation of the phase that
enters the first-order term of the asymptotic expression, as well as
variations introduced by a departure from the asymptotic expression,
in particular by
acoustic glitches in the outer convective envelope. 
We tested these two formulations on a standard solar model, for which
the $l=1$ modes are not mixed and, therefore, are known a priori. The
results of that test confirm that Eq.~(\ref{omega_a2}) reproduces 
the true $l=1$ frequencies significantly better (the details are discussed in Appendix~\ref{apb}).

    \begin{figure*} 
         \begin{minipage}{1.\linewidth}  
               \rotatebox{0}{\includegraphics[width=0.48\linewidth]{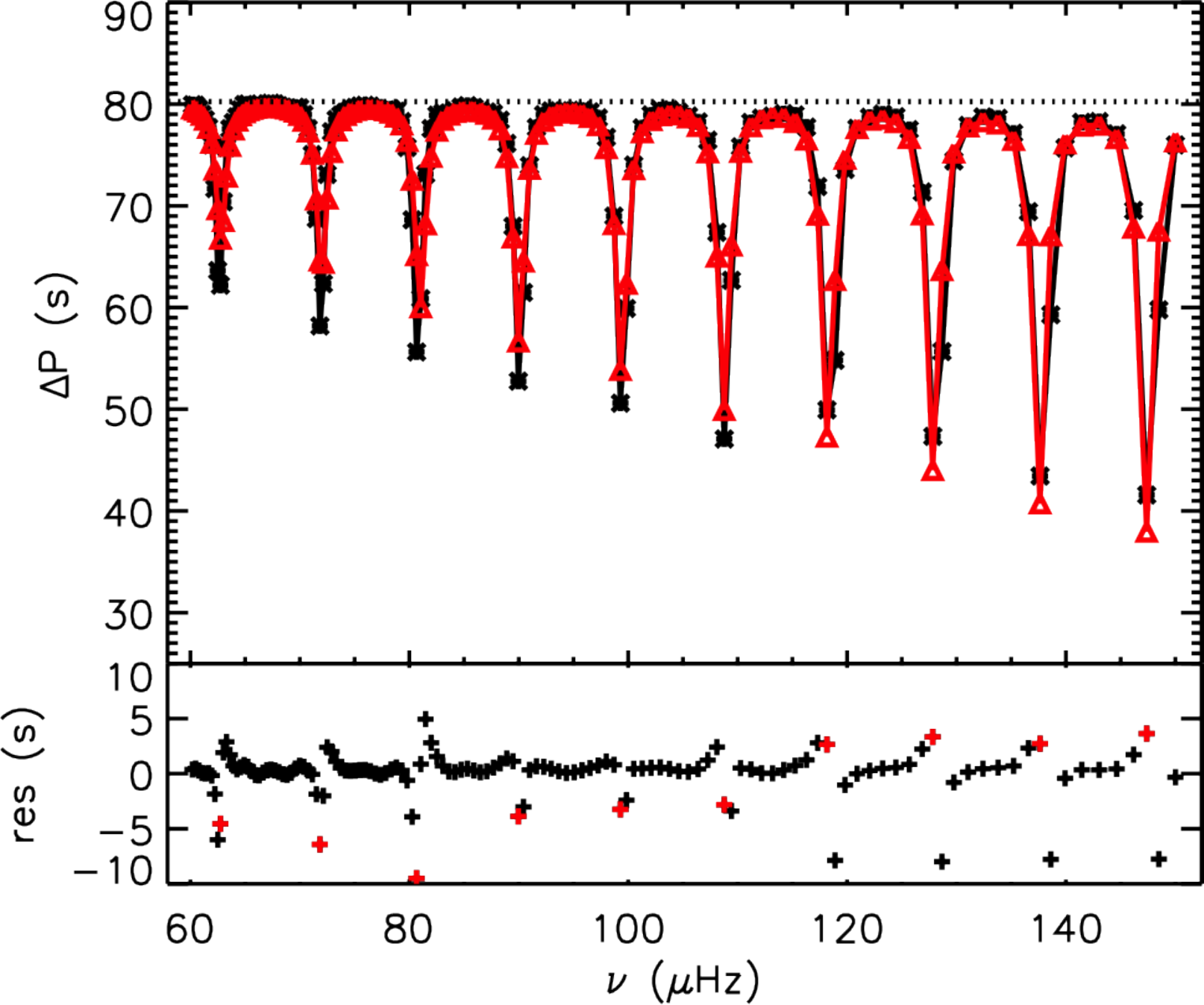}} \hspace{0.5cm}
\rotatebox{0}{\includegraphics[width=0.48\linewidth]{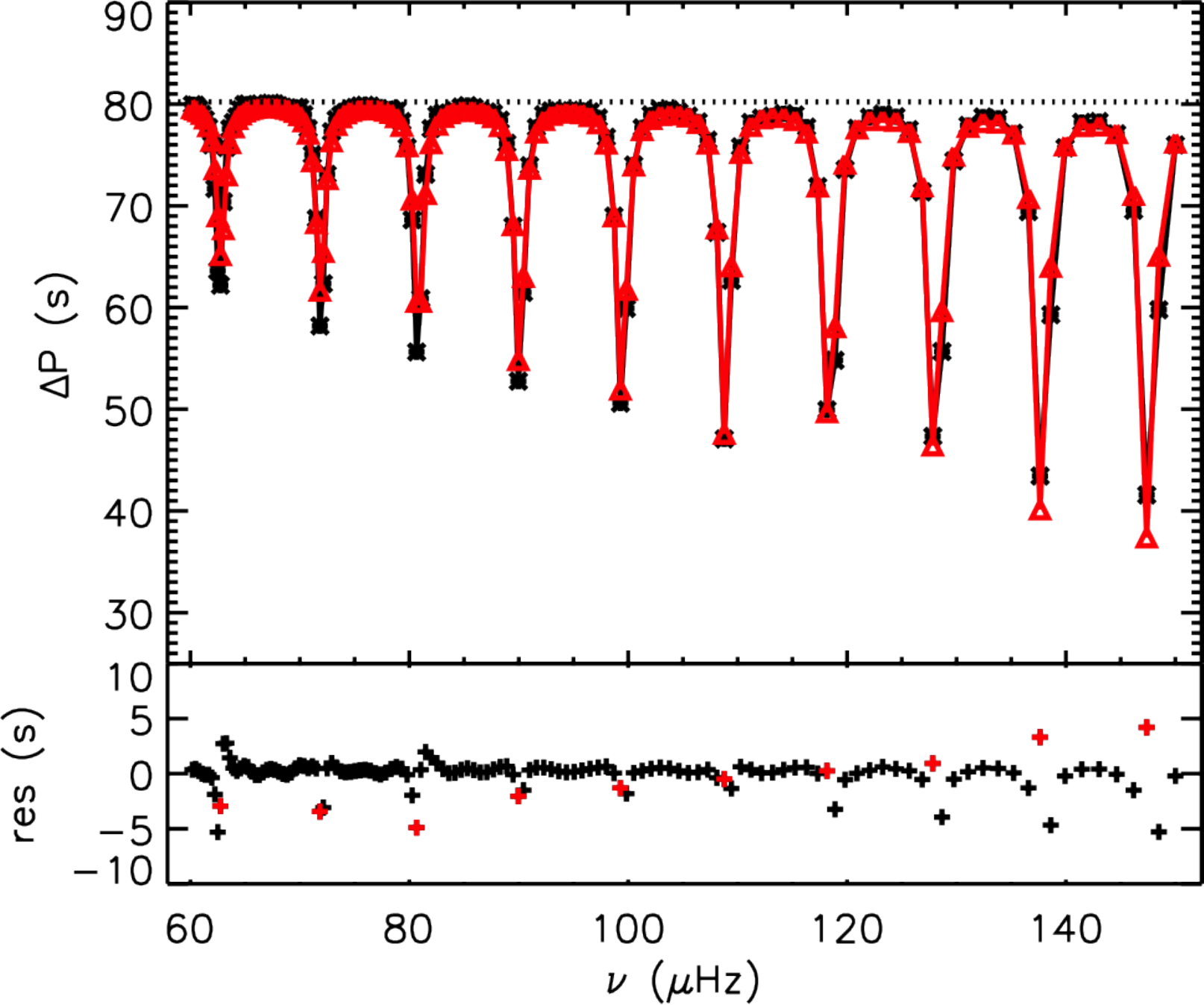}}
         \end{minipage}
\caption{Top panels: comparison between the period spacing derived from ADIPLS for
our RGB-1 model (with no core glitch)  (black line and asterisks) and that
obtained from Eq.~(\ref{ps_coupling_2})  with the most likely
parameters from our fit,  performed in the frequency range shown in the figures (red line and triangles).  Left is for 
$\omega_{{\rm a},n}$ estimated through eq.~(\ref{omega_a1}) and right is for 
$\omega_{{\rm a},n}$  estimated through eq.~(\ref{omega_a2}).  Bottom panels: the residuals (`ADIPLS period spacing' $-$ `analytical period spacing') for each case. The red symbols mark the residuals at the minima of the ADIPLS period spacing.}
   \label{caso2_best_c}
       \end{figure*}

Figure~\ref{caso2_best_c} shows the comparison between the period spacing
computed from the ADIPLS results for a glitch-less RGB  model and those obtained from
Eq.~(\ref{ps_coupling_2}) adopting the parameters of the most likely
solution found from the fit. The left and right panels differ only
in the option adopted for the estimate of $\omega_{{\rm a},n}$ (eqs~(\ref{omega_a1}) and (\ref{omega_a2}), respectively). While the
quality of the fit for option 2 is the better of the two, it is quite clear from Fig.~\ref{caso2_best_c} that for 
both options the analytic formulation fails to reproduce the
dips. This is not entirely surprising, because models predict that the coupling
coefficient for stars ascending the RGB  should be frequency
dependent \citep{jiang14,hekker18}. That dependence, which is present for models with $\nu_{\rm max}\lesssim 100~\mu$Hz, results from the fact that the
acoustic cavity becomes deeper with increasing frequency while the
g-mode cavity barely changes, resulting in a decrease
of the width of the evanescent region, with increasing frequency. According to
\cite{jiang14}, the frequency dependence of $q$
is well represented by a linear function, for models ascending the RGB.

To test the frequency dependence of $q$, we have performed a third fit to the period spacings
obtained from ADIPLS, considering a linear frequency dependent coupling
coefficient, defined by
\begin{equation}
\label{q_freq}
q=q_1\left[\alpha\left(\nu/\nu_{\rm max}-1\right)+1\right],
\label{q}
\end{equation}
thus replacing the parameter $q$ by the pair of parameters
($q_1,\alpha$).
Since $q$ enters the definition of the coupling phase  $\varphi$, its
dependence on frequency needs to be taken into
account when differentiating $\varphi$ in Eq.~(\ref{ps_coupling}). In that case,
the analytical expression for the period spacing in the presence of
mode coupling, without a buoyancy glitch,
previously given by Eq.~(\ref{ps_coupling_2}), is replaced by
\begin{eqnarray}
\label{ps_coupling_3}
{{\frac{\Delta P}{\Delta P_{\rm as}}\Resize{7.5cm}{\approx
      \hspace{-0.08cm}\left[1+{
          {\frac{\omega^2}{\omega_{\rm g}}\frac{q}{\omega_{\rm p}}\left[\sin^2\left(\frac{\omega-\omega_{{\rm a},n}}{\omega_{\rm p}}\right)+q^2\cos^2\left(\frac{\omega-\omega_{{\rm a},n}}{\omega_{\rm p}}\right)\right]^{-1}}}+Q\left(\omega\right)\right]^{-1}}}}
\hspace{-0.2cm},\hspace{-0.1cm}
\end{eqnarray}
where the function $Q(\omega)$ is given by,
\begin{equation}
Q(\omega)=\frac{q_1\alpha\omega^2}{2\pi\nu_{\rm max}\omega_{\rm g}}\left[q^2\cot\left(\frac{\omega-\omega_{{\rm a},n}}{\omega_{\rm p}}\right)+\tan\left(\frac{\omega-\omega_{{\rm a},n}}{\omega_{\rm p}}\right)\right]^{-1}.
\end{equation}

\begin{table*}
	\label{parameters_ng}
	\caption{Parameters derived from the fit of the analytical expression in
		Eq.~(\ref{ps_coupling_3})  to the 
		period spacing derived from ADIPLS for the model RGB-1. Their distributions are shown in Fig.~\ref{caso2_qfreq}. The values shown correspond to the median of the distributions and the $68\%$ confidence intervals. We recall that $\Delta\nu_{\rm as}\approx\omega_{\rm p}/2$.}
	\begin{minipage}{0.7\textwidth}
		\resizebox{\linewidth}{!}{%
			\begin{tabular}{|c|c|c|c|c|}
				\hline
				$\Delta P_{\rm as}$ (s)  & $\omega_{\rm p}/2$ $(\mu{\rm Hz})$ & $q_1$ & $C$ $(\mu{\rm Hz}^2)$ & $\alpha$\\
				\hline
				& & & & \\
				$80.10_{-0.09}^{+0.10}$& $11.02_{-1.17}^{+1.55}$& $0.128_{-0.015}^{+0.015}$& $19.93_{-0.53}^{+0.53}$ & $0.692_{-0.048}^{+0.046}$ \\
				& & & & \\
				\hline
			\end{tabular}
		}
	\end{minipage}
\end{table*}

\begin{figure*} 
         \begin{minipage}{1.\linewidth}  
               \rotatebox{0}{\includegraphics[width=1.0\linewidth]{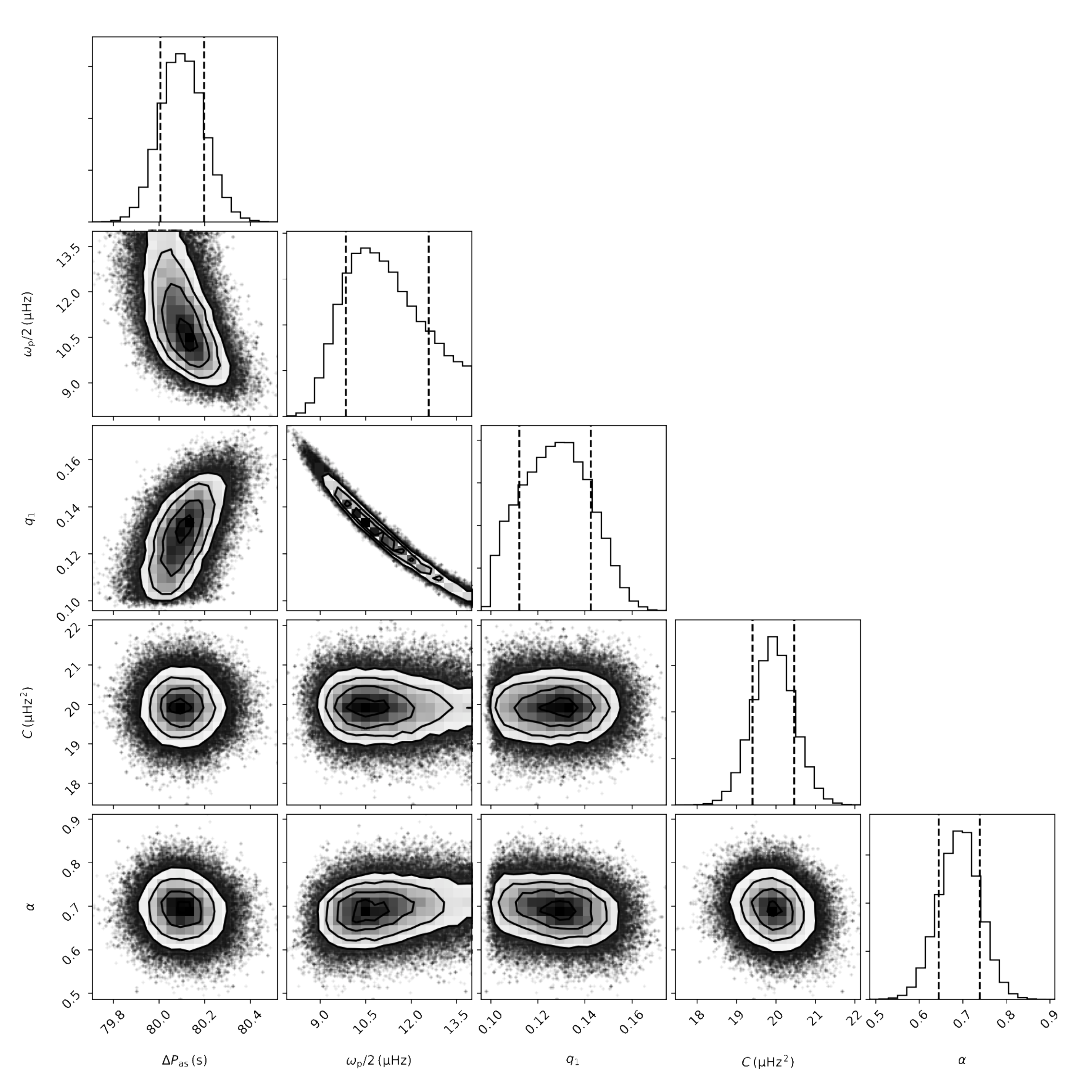}}
         \end{minipage}
\caption{Marginalised distributions for the parameters considered in
  the fit of the rhs of Eq.~(\ref{ps_coupling_3}) to the period
  spacing derived from ADIPLS for the RGB model below the luminosity
  bump (model RGB-1, with no glitch), when considering a frequency dependent $q$
  according to Eq.~(\ref{q_freq}).}
   \label{caso2_qfreq}
       \end{figure*}

\begin{figure} 
         \begin{minipage}{1.\linewidth}  
\rotatebox{0}{\includegraphics[width=1.0\linewidth]{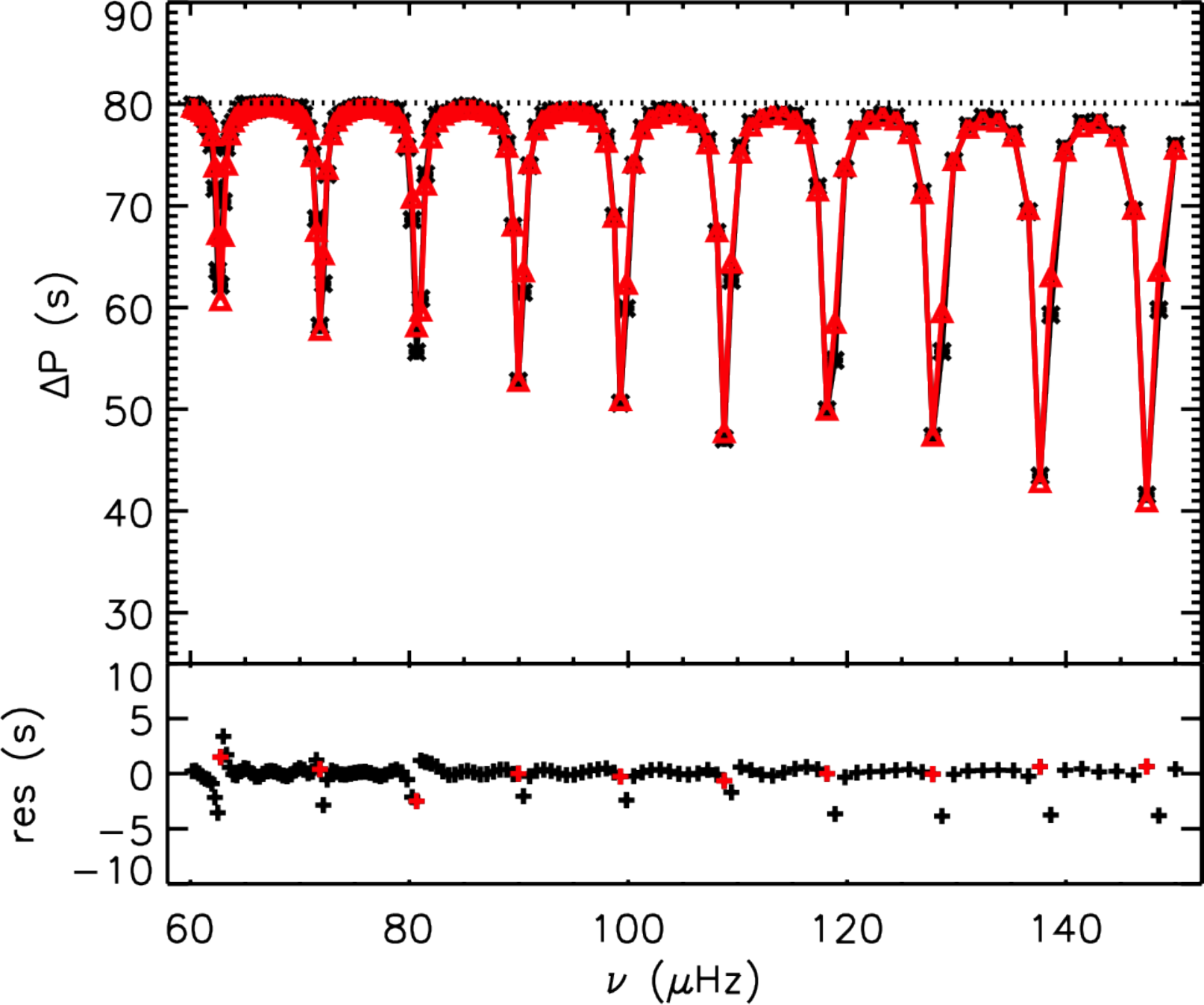}}
         \end{minipage}
\caption{Top panel: Comparison between the period spacing derived from ADIPLS
 for the RGB-1 model (with no core glitch)  (black line and asterisks) and that obtained from
Eq.~(\ref{ps_coupling_3})  with a frequency dependent $q$, with the
most likely parameters from our fit,  performed in the frequency range shown in the figure (red line and triangles). Bottom panel: the residuals (`ADIPLS period spacing' $-$ `analytical period spacing'). The red symbols mark the residuals at the minima of the ADIPLS period spacing.}
   \label{caso2_best_qfreq}
       \end{figure}

The results of the fit when $q$ is considered frequency dependent are shown in Figs~\ref{caso2_qfreq} and
\ref{caso2_best_qfreq}. Clearly,  the parameter
$\alpha$ introduced in association to the frequency dependence of $q$ is well
constrained in a region that excludes zero (Fig.~\ref{caso2_qfreq}), confirming
that a frequency-independent $q$ does not provide a good fit. The quality of
the fit is also found to be substantially better than when $q$ is
taken to be constant, a fact that is
noticeable when
comparing Fig.~\ref{caso2_best_qfreq} with
Fig.~\ref{caso2_best_c}. Naturally, the larger the number of radial orders fitted, the more noticeable the frequency dependence becomes. When fitting observational data, with a limited number of radial orders available, one must thus verify whether adding one additional parameter to characterise the frequency dependence of $q$ is a relevant option. 

The increase of the depth of the acoustic cavity with frequency, used as an argument
to make $q$ frequency dependent,
influences also the parameter
$\omega_{\rm p}$, defined by Eq.~(\ref{omega_p}), which, as a result, decreases with frequency. Due to the
roles of $q$ and $\omega_{\rm p}$ in Eq.~(\ref{ps_coupling_2}), this 
frequency dependence of $\omega_{\rm p}$, not accounted for in the previous fit, would emphasise even further the
need for a frequency dependence of $q$. However,  according to the work by \cite{jiang14}, the frequency
dependance of $\omega_{\rm p}$ is generally small and becomes even smaller at the highest
frequencies, being well fitted by a second order polynomial. To verify
whether this dependence influences the
significance found in our previous fit for a frequency dependence of $q$, we have
performed a fourth fit, taking both $q$ and $\omega_{\rm p}$ as frequency
dependent, allowing that dependence to go to second order. We found
that the parameters characterising the frequency dependence of
$\omega_{\rm p}$ were consistent with this parameter being constant,
confirming that its frequency dependence is small. Moreover, the $q$
parameter was confirmed to be well described by a linear dependence on
frequency (the second order term being
consistent with zero), with an $\alpha$ value consistent with that
found in the previous fit, albeit less well constrained, as expected,
given the larger number of parameters being fitted.

\section{Combined effect of mode coupling and a buoyancy glitch}
\label{g_c}

When both mode coupling and a structural glitch in the core are present, the exact
form of Eq.~(\ref{ps_coupling_glitch}) depends again on the functional
form adopted to model the glitch. In \cite{cunha15} the authors
presented the expression for the case of a core glitch placed in the
outer half of the g-mode cavity, where the glitch was modelled with a
Dirac delta function \footnote{We note again that there is a typo in that
  expression and the reader is advised to see footnote 1 in  the present
  paper for further details.} and the coupling coefficient $q$ was
considered to be independent of frequency.

Here, we provide the analytical expression for the case of a
glitch in the outer half of the cavity\footnote{In the case of mixed modes, the signature from the Gaussian-like glitch is not invariant with respect to symmetric changes about the center of the g-mode cavity ({\it cf.} discussion in Appendix~\ref{apa}).} modelled by the Gaussian-like
function discussed in Sec.~\ref{g}. We thus assure that the functional form adopted for the
glitch represents adequately the core
glitch seen in the RGB model located
at the luminosity bump (model RGB-2), discussed in Sec.~\ref{g}. In the case of a
Gaussian-Like glitch, Eq.~(\ref{ps_coupling_glitch}) can be
re-written as (see Appendix~\ref{apa}, for details),
\begin{equation}
\label{F_c,g}
\frac{\Delta P}{\Delta P_{\rm as}}\approx\left[1-\mathcal{F}_{G,C}\right]^{-1}
\end{equation}
where,
{\begin{eqnarray}
\hspace{-0.cm}\mF_{\rm {G,C}}\hspace{-0.0cm} &= &\\
  &&\scriptsize{\hspace{-1.6cm}\frac{\omega^2}{\omega_{\rm
    g}}\frac{\rmd\varphi}{\rmd\omega}\left\{1+\frac{A_{\rm G}
    f_{\omega}^{\Delta_{\rm g}}}{B^2}\left[\cos\left(\beta_{1,\varphi}\right)-A_{\rm G}
    f_{\omega}^{\Delta_{\rm g}} \sin^2\left(\beta_{2,\varphi}\right)\right]\right\}\nonumber} \\ 
 &&\scriptsize{\hspace{-1.7cm} -\hspace{-0.0cm} \frac{\omega_{\rm g}^\star}{\omega_{\rm g}}\hspace{-0.05cm} \frac{
    A_{\rm G}
    f_{\omega}^{\Delta_{\rm g}}}{B^2}\hspace{-0.1cm}  \left\{\cos\left(\beta_{1,\varphi}\right)\hspace{-0.1cm} +\hspace{-0.1cm} \left[\omega/\omega_{\rm g}^\star(1-4\Delta_{\rm g}^2/\omega^2)-
      A_{\rm G}
        f_{\omega}^{\Delta_{\rm g}}\right]\hspace{-0.1cm} \sin^2\left(\beta_{2,\varphi}\right)\right\}}\nonumber
\label{fgc}
\end{eqnarray}
\normalsize
and $B^2$ is given by 
\begin{eqnarray}
B^2= {\left[1-0.5 A_{\rm G}
      f_{\omega}^{\Delta_{\rm g}}\cos\left(\beta_{1,\varphi}\right)\right]^2+\left[A_{\rm G} f_{\omega}^{\Delta_{\rm g}}\sin^2\left(\beta_{2,\varphi}\right)\right]^2}.
\label{b2gc}
\end{eqnarray}}
\normalsize
Here, the arguments of the sinusoidal functions are changed
with respect to the Gaussian glitch case presented in Sec.~\ref{g},
now being given by $\beta_{1,\varphi}=\beta_{1}+2\varphi$ and
$\beta_{2,\varphi}=\beta_2+\varphi$, where $\beta_1$, $\beta_2$ and other
glitch-related quantities are defined in that section. This change is a consequence of the dependence of
the glitch phase on the coupling phase. Moreover, based on the results of
Sec.~\ref{c}, when testing this analytical expression against model
data, $q$ will be taken to depend on the
frequency according to Eq.~(\ref{q}).  The explicit form of
$\rmd\varphi/\rmd\omega$ is obtained from the analytical differentiation of Eq.~(\ref{varphi}).

To test the analytical expression defined by
Eqs~(\ref{F_c,g})-(\ref{b2gc}) we fit it to the period spacings derived
from the ADIPLS frequencies for our RGB model located at the luminosity
bump (model RGB-2), following the same
approach as in Sec.~\ref{g}. We start by fixing the glitch parameters, $A_{\rm G}$, $\Delta_{\rm g}$, $\omega_{\rm g}^\star$,
and $\delta$ at the values derived in Sec.~\ref{g} and adopt
Eq.~(\ref{omega_a2}) to describe the pure acoustic frequencies
$\omega_{{\rm a},n}$. The problem thus involves fitting five parameters: two
characterising the global seismic properties ($\omega_{\rm p}$ and $\Delta P_{\rm as}$), two other characterising
the mode coupling ($q_1$ and $\alpha$), and one characterising the relation between the
frequencies of pure acoustic modes ($C$). 

\begin{figure*} 
         \begin{minipage}{1.\linewidth}  
               \rotatebox{0}{\includegraphics[width=0.48\linewidth]{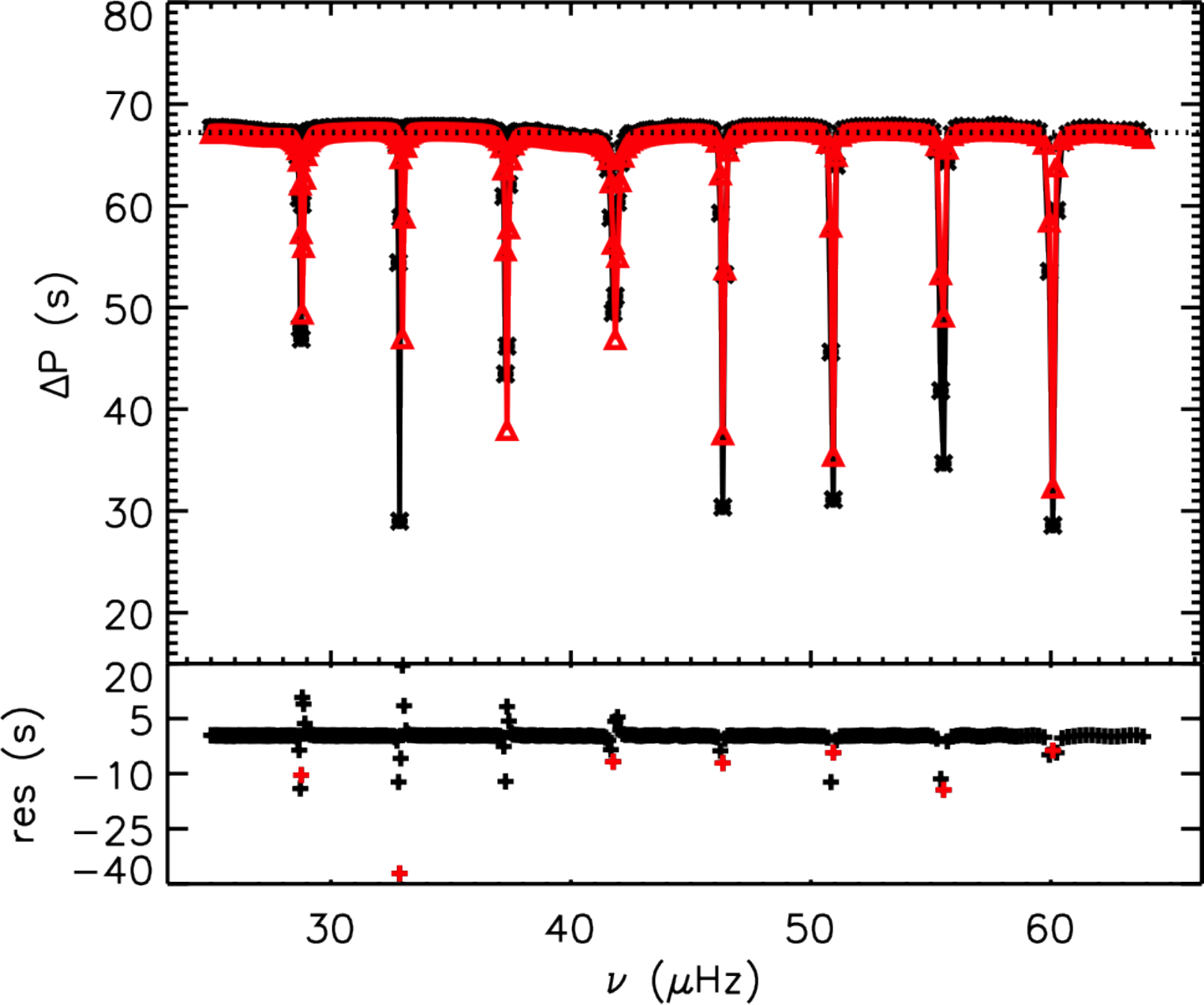}}
                         \hspace{0.5cm}
               \rotatebox{0}{\includegraphics[width=0.48\linewidth]{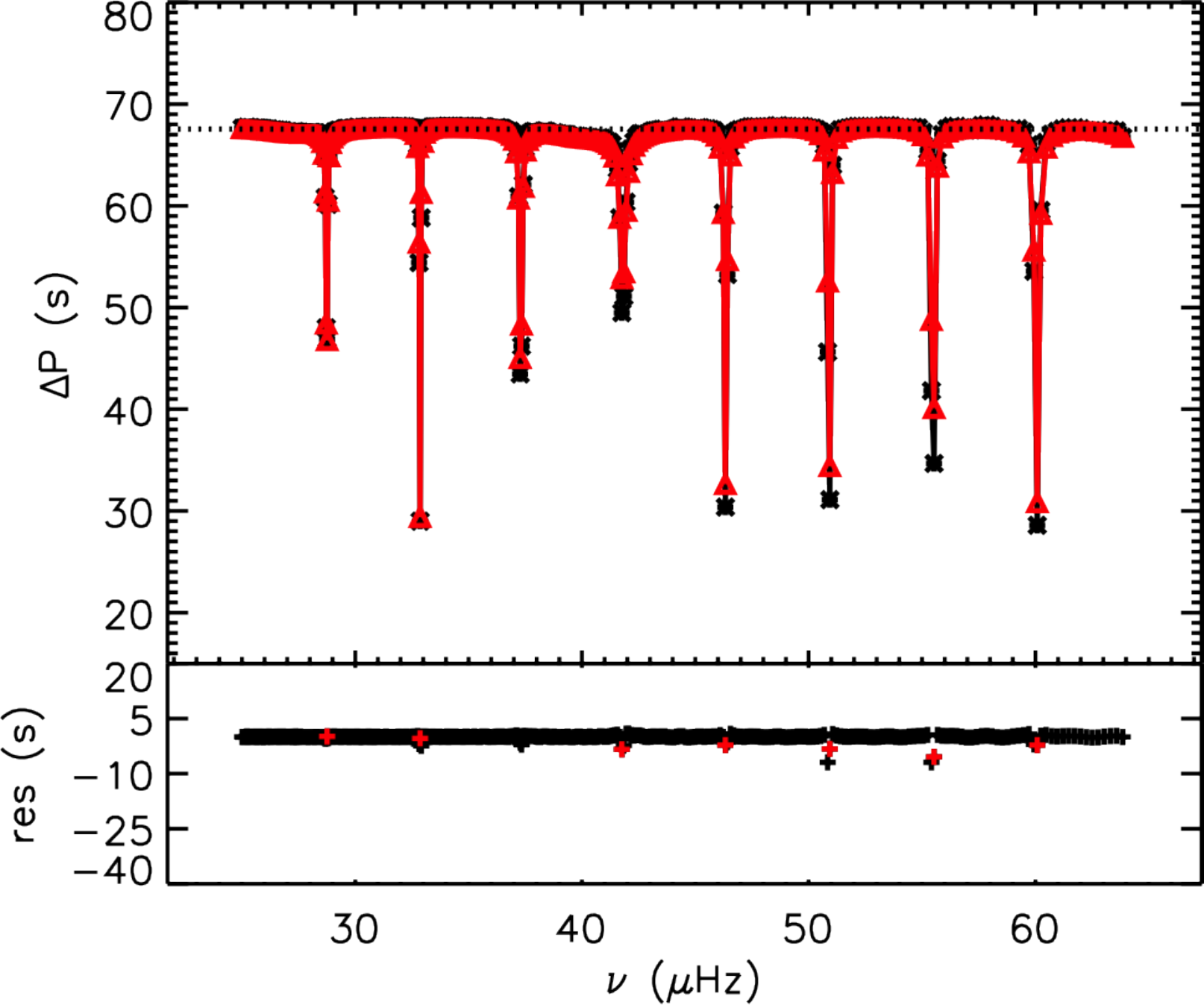}}
               
         \end{minipage}
\caption{Top panels: comparison between the period spacing derived from ADIPLS
 for our RGB-2 model (with a core glitch)  (black line and asterisks) with that obtained from
Eqs~(\ref{F_c,g})-(\ref{b2gc}) with the most likely parameters from our fit,  performed in the frequency range shown in the figure (red line and triangles). Left is for $\omega_{{\rm a},n}$ derived from Eq.~(\ref{omega_a2}) and right for $\omega_{{\rm a},n}$  left as free parameters in the fit. Moreover, $q$ was taken to depend
linearly on the frequency, according to eq.~(\ref{q}) and the glitch parameters were fixed from
the outset, based on the results of section~\ref{g}. Bottom panels: the residuals (`ADIPLS period spacing' $-$ `analytical period spacing'). The red symbols mark the residuals at the minima of the ADIPLS period spacings.
}
   \label{caso3_Cfixed_qvar}
       \end{figure*}

The results of the fit are shown in
Fig.~\ref{caso3_Cfixed_qvar}, left panel, where it can readily be noticed that
the model, with the parameters from the best fit, fails to reproduce
adequately the ADIPLS results near some of the coupling dips. We note,
however, that in the model under consideration the variation of the
period spacing near the pure acoustic frequencies is extremely large. Hence,
the accuracy to which one derives the  $\omega_{{\rm a},n}$  values can be of
importance to the quality of the fit. In particular, it is
important to establish whether the failure to properly fit the model
period spacings is a consequence of the inadequacy of the
analytical expression used in the fit, or if it results, instead,
from the insufficient accuracy of the
pure acoustic frequencies estimated from  Eq.~(\ref{omega_a2}). 

In the case of the Sun, discussed in Appendix~\ref{apb}, we find that some frequency-dependent residuals remain
when the $l=1$ model frequencies are compared with the estimates
obtained from Eqs.~(\ref{omega_a1}) and (\ref{omega_a2}).  To test the
impact of small variations of $\omega_{{\rm a},n}$ on the fit, we performed
the fit on RGB-2 model again, under two different conditions: i) estimating $\omega_{{\rm a},n}$
from Eq.~(\ref{omega_a1}) and, ii) letting the $\omega_{{\rm a},n}$ values be
independent free parameters. In the first case we found that the
quality of the fit got worse compared to Fig.~\ref{caso3_Cfixed_qvar} (left panel), reflecting that the estimates of the pure acoustic
frequencies worsened, as expected from the results for the solar
case.
On the other hand, the fit was much improved when these
frequencies were let free, as seen from the inspection
of Fig.~\ref{caso3_Cfixed_qvar}, right panel.

\begin{figure} 
         \begin{minipage}{1.\linewidth}  
               \rotatebox{0}{\includegraphics[width=1.0\linewidth]{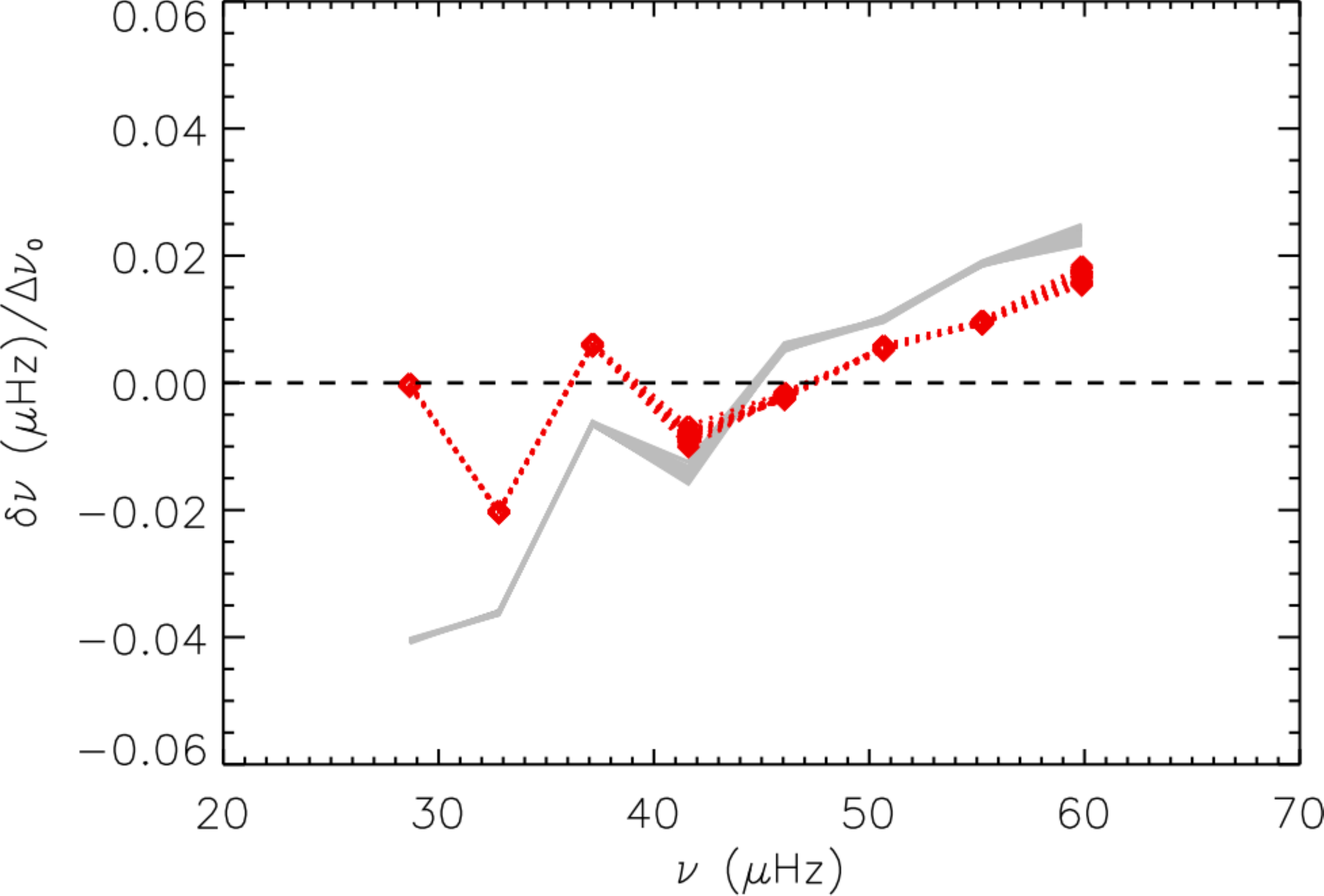}}
         \end{minipage}
\caption{Difference between the pure acoustic frequencies
  $\nu_{a,n}=\omega_{{\rm a},n}/2\pi$ estimated from fitting
  Eqs~(\ref{F_c,g})-(\ref{b2gc}) to the model period spacings with $\omega_{{\rm a},n}$ taken as
  free parameters and those estimated from otherwise similar fits but
  with $\omega_{{\rm a},n}$ given by: i) Eq.~(\ref{omega_a1})
  (grey curve) and ii) Eq.~(\ref{omega_a2}) (red-dashed curve; diamonds). The difference is scaled by the large frequency separation and is shown
  for a sample of 30 best fitted models (with similar likelihood),
  whose superposition is reflected in the slight broadening of the
  lines plotted.
}
   \label{dif_wa}
       \end{figure}

 The significant improvement in the fit
observed in the last case discussed above is not a surprise in itself, given
the increase in the number of free parameters. It is, thus, important
to assess whether the set of $\omega_{{\rm a},n}$ retrieved from the fit in this
case makes physical sense, or, rather, is simply a combination of unrelated
departures from the previous estimates that results from the fitting
procedure attempting to correct  a possible inadequacy of the analytical
representation of the period spacing. To clarify this
matter we computed the differences between the frequencies
$\omega_{{\rm a},n}$ obtained from the fit where these have been left as
free parameters and those obtained from Eqs~(\ref{omega_a1}) and (\ref{omega_a2}) with values of $C$ derived from the corresponding fits. The
comparison is shown in Fig.~\ref{dif_wa} for a sample of 30 best fitted models (with similar likelihood). The comparison of the results in Fig.~\ref{dif_wa}
with those found for the solar model S (Fig.~\ref{comp_solar}), is
very encouraging.  In particular, the difference between the
freely-determined $\omega_{{\rm a},n}$ and the $\omega_{{\rm a},n}$ estimated from
Eq.~(\ref{omega_a1}) shows a trend with frequency that resembles that found for the solar
model when the frequencies estimated by Eq.~(\ref{omega_a1}) are subtracted
from the exact $l=1$ model frequencies. This is particularly significant, because the estimate of 
$\omega_{{\rm a},n}$ through Eq.~(\ref{omega_a1}) does not depend on any
interpolation procedure whose adequacy may be different for the Sun
and for a red-giant model. We note that 
the differences shown in Fig.~\ref{dif_wa} (scaled to the model  large frequency
separation) for the RGB model are about one order of magnitude larger than those found for the solar model, for the same range of radial orders around
$\nu_{\rm max}$. This is true both for the differences illustrated by the curve in grey and for those illustrated by the curve in red. That can be understood from the fact that 
both the
frequency signature of acoustic glitches associated with the helium second
ionization and the large-scale frequency variation associated to surface effects are,
after scaling by the large separation, about one order of
magnitude larger in the RGB model compared to the solar model
\citep[e.g.][]{broomhall14,Houdek07}. In addition, in the solar model the signature of the glitches on the frequencies is better resolved, because of the denser acoustic frequency spectrum. These two facts explain that the difference between the true
frequencies and the estimated ones is more 
significant in the RGB model. Given the evidence above, we are confident that the
estimates of the $\omega_{{\rm a},n}$ derived from the fit of the analytical expression defined by
Eqs~(\ref{F_c,g})-(\ref{b2gc}) to the 
period spacing of this luminous RGB model, are the best of the three
possible estimates considered here. So, we trust that having $\omega_{{\rm a},n}$ free
when fitting such a luminous RGB star is the best option in the present case.

Following on the results discussed above, we attempted to perform the
same fits, with the different options for the pure acoustic
frequencies,  by considering the glitch parameters to be
free. Unfortunately, none of the options considered produced
reasonable contraints to the glitch parameters. This is because the
$\chi^2$ minimisation is heavily influenced by small
departures between the analytical expression and the model data at the frequencies around the acoustic
dips, where the period spacing varies abruptly. Hence, the quality of the fit is not
sufficiently sensitive to the glitch parameters in this case. As an example, we
find that fits
corresponding to a no-glitch solution (with a flat period spacing
everywhere except around the coupling dips) can have a similar $\chi^2$
as that shown in Fig.~\ref{caso3_Cfixed_qvar}, right panel. This experience points
towards the need to adopt a different strategy, perhaps not based
on a global $\chi^2$ criteria, to constrain the parameters of the
glitch from the analytical expression presented in
Eqs~(\ref{F_c,g})-(\ref{b2gc}). However, the
results found when fixing the parameters of the glitch do confirm that the proposed
analytical expression provides a good representation of the model period
spacing in the presence of a core structural glitch and mode coupling.

\section{Conclusions}
\label{conclusion}

In this work we have tested an analytical representation of the dipole mode period spacing derived from asymptotic analysis against model data. The analytical expression is relevant for the modelling of stars exhibiting pure gravity modes as well as stars exhibiting mixed modes. The impact of different types of structural glitches that may be present in the cores of the stars has been fully accounted for, as has the coupling between p and g modes, when present. Rotation effects have not been considered.  

{With the exception of the amplitude in one of the cases considered, our results show that the buoyancy-glitch parameters can be adequately recovered by fitting the proposed analytical expression to model data consisting of pure gravity modes.} We stress that unlike in previous works, the analytical expression tested here is valid also when the glitch is not small and, consequently, when the glitch-induced period spacing variations are not sinusoidal. This is important because when the glitch is not small the period spacing is asymmetric with respect to the asymptotic value and fitting it with a sinusoidal function, such as that predicted in the small glitch limit, may lead to a biased estimation of the asymptotic period spacing, as well as of the glitch parameters. 

{For the case of pure gravity modes, the relative differences between the glitch parameters estimated directly from the buoyancy frequency and those inferred from the fits of the analytical expression to the period spacings for the two cases studied here are smaller than 7$\%$, for all parameters, but the amplitudes. For the step-like glitch, the amplitude value estimated from the buoyancy frequency is 11$\%$ larger than the median of the distribution inferred from the fit to the period spacings. However,  Fig.~\ref{fig:glitch}\,b reassures us that the glitch amplitude is adequately recovered when considering the uncertainty introduced by the adopted step-function model.  In  the case of the glitch modelled by a Gaussian, the inferred amplitude is found to be about 60$\%$ larger than expected. This difference is likely related to our inability to correctly model the eigenfunction inside the glitch, where the asymptotic analysis fails. Further tests shall be performed in future work covering a larger set of models, to calibrate the inferred amplitude against the true one and establish the range of applicability of the expression in the case of the Gaussian-like glitch.}

In addition, we find that the analytical expression describing the period spacing for mixed modes propagating in the presence of a buoyancy glitch represents well the period spacing derived numerically for a red-giant model exhibiting such a glitch. However, our results indicate that in the case considered here, of a RGB star at the luminosity bump, the fit of the analytical expression to the model period spacing based on a global $\chi^2$ minimisation criteria does not allow us to contrain the glitch parameters. This is because the period spacing variations are dominated by the effect of the mode coupling. Alternative approaches to fit  the analytical expression to  model data that may allow to highlight the impact of the glitch on the oscillation periods and, thus, constrain the glitch parameters, are being considered and will be discussed in a future work. 

Finally, our fit of the analytical expression to mixed-mode model data in the absence of buoyancy glitches indicates a clear frequency dependence of the coupling coefficient $q$. This dependance, which is theoretically expected for stars with $\nu_{\rm max}$ smaller than $\sim 100\mu$Hz, may need to be considered when fitting data of intermediate to high luminosity red-giant stars, depending on the number of radial orders observed. 

Interestingly, our results also show that by fitting the proposed analytical expression to the dipole mixed-mode period spacing, it might be possible to extract the frequencies of the pure acoustic dipole modes that would exist, had these modes not been mixed in red-giant stars.

\section*{Acknowledgements}
{We thank the referee, S\'ebastien Deheuvels, for making very pertinent comments on the first version of this manuscript. We also thank Stefano Garcia for useful comments and for providing a modified RGB model with a glitch of higher amplitude to test the results of Appendix A under different conditions.}  This work was supported by FCT - Funda\c c\~ao para a Ci\^encia e a Tecnologia  through national funds and by FEDER through COMPETE2020 - Programa Operacional Competitividade e Internacionaliza\c c\~ao by these grants: UID/FIS/04434/2013 \& POCI-01-0145-FEDER-007672, PTDC/FIS-AST/30389/2017 \& POCI-01-0145-FEDER-030389. Funding for the Stellar Astrophysics Centre is provided by The Danish National Research Foundation (Grant DNRF106). The research was supported by the ASTERISK project (ASTERoseismic Investigations with SONG and Kepler) funded by the European Research Council (Grant agreement no.: 267864). D.S is the recipient of an Australian Research Council Future Fellowship (project number FT1400147).




\bibliographystyle{mnras}
\bibliography{solar-like_v0} 

\begin{thebibliography}{}
\makeatletter
\relax
\def\mn@urlcharsother{\let\do\@makeother \do\$\do\&\do\#\do\^\do\_\do\%\do\~}
\def\mn@doi{\begingroup\mn@urlcharsother \@ifnextchar [ {\mn@doi@}
  {\mn@doi@[]}}
\def\mn@doi@[#1]#2{\def\@tempa{#1}\ifx\@tempa\@empty \href
  {http://dx.doi.org/#2} {doi:#2}\else \href {http://dx.doi.org/#2} {#1}\fi
  \endgroup}
\def\mn@eprint#1#2{\mn@eprint@#1:#2::\@nil}
\def\mn@eprint@arXiv#1{\href {http://arxiv.org/abs/#1} {{\tt arXiv:#1}}}
\def\mn@eprint@dblp#1{\href {http://dblp.uni-trier.de/rec/bibtex/#1.xml}
  {dblp:#1}}
\def\mn@eprint@#1:#2:#3:#4\@nil{\def\@tempa {#1}\def\@tempb {#2}\def\@tempc
  {#3}\ifx \@tempc \@empty \let \@tempc \@tempb \let \@tempb \@tempa \fi \ifx
  \@tempb \@empty \def\@tempb {arXiv}\fi \@ifundefined
  {mn@eprint@\@tempb}{\@tempb:\@tempc}{\expandafter \expandafter \csname
  mn@eprint@\@tempb\endcsname \expandafter{\@tempc}}}

\bibitem[\protect\citeauthoryear{{Aerts}, {Christensen-Dalsgaard}  \&
  {Kurtz}}{{Aerts} et~al.}{2010}]{aerts10}
{Aerts} C.,  {Christensen-Dalsgaard} J.,   {Kurtz} D.~W.,  2010,
  {Asteroseismology}

\bibitem[\protect\citeauthoryear{{Aerts}, {Mathis}  \& {Rogers}}{{Aerts}
  et~al.}{2018}]{aertsetal18}
{Aerts} C.,  {Mathis} S.,   {Rogers} T.,  2018, arXiv e-prints, \href
  {http://adsabs.harvard.edu/abs/2018arXiv180907779A} {}

\bibitem[\protect\citeauthoryear{{Baglin}, {Auvergne}, {Barge}, {Deleuil},
  {Catala}, {Michel}, {Weiss}  \& {COROT Team}}{{Baglin}
  et~al.}{2006}]{baglin06}
{Baglin} A.,  {Auvergne} M.,  {Barge} P.,  {Deleuil} M.,  {Catala} C.,
  {Michel} E.,  {Weiss} W.,   {COROT Team} 2006, in {Fridlund} M.,  {Baglin}
  A.,  {Lochard} J.,   {Conroy} L.,  eds,  ESA Special Publication Vol. 1306,
  The CoRoT Mission Pre-Launch Status - Stellar Seismology and Planet Finding.
  p.~33

\bibitem[\protect\citeauthoryear{{Bossini} et~al.,}{{Bossini}
  et~al.}{2015}]{bossini15}
{Bossini} D.,  et~al., 2015, \mn@doi [\mnras] {10.1093/mnras/stv1738}, \href
  {http://adsabs.harvard.edu/abs/2015MNRAS.453.2290B} {453, 2290}

\bibitem[\protect\citeauthoryear{{Brassard}, {Fontaine}, {Wesemael}  \&
  {Hansen}}{{Brassard} et~al.}{1992}]{brassard92}
{Brassard} P.,  {Fontaine} G.,  {Wesemael} F.,   {Hansen} C.~J.,  1992, \mn@doi
  [\apjs] {10.1086/191668}, \href
  {http://adsabs.harvard.edu/abs/1992ApJS...80..369B} {80, 369}

\bibitem[\protect\citeauthoryear{{Broomhall} et~al.,}{{Broomhall}
  et~al.}{2014}]{broomhall14}
{Broomhall} A.-M.,  et~al., 2014, \mn@doi [\mnras] {10.1093/mnras/stu393},
  \href {http://cdsads.u-strasbg.fr/abs/2014MNRAS.440.1828B} {440, 1828}

\bibitem[\protect\citeauthoryear{{Brown}}{{Brown}}{1991}]{brown91}
{Brown} T.~M.,  1991, \mn@doi [\apj] {10.1086/169900}, \href
  {http://adsabs.harvard.edu/abs/1991ApJ...371..396B} {371, 396}

\bibitem[\protect\citeauthoryear{{Christensen-Dalsgaard}}{{Christensen-Dalsgaard}}{2008a}]{jcd08a}
{Christensen-Dalsgaard} J.,  2008a, \mn@doi [\apss]
  {10.1007/s10509-007-9675-5}, \href
  {http://adsabs.harvard.edu/abs/2008Ap%26SS.316...13C} {316, 13}

\bibitem[\protect\citeauthoryear{{Christensen-Dalsgaard}}{{Christensen-Dalsgaard}}{2008b}]{jcd08b}
{Christensen-Dalsgaard} J.,  2008b, \mn@doi [\apss]
  {10.1007/s10509-007-9689-z}, 316, 113

\bibitem[\protect\citeauthoryear{{Christensen-Dalsgaard}}{{Christensen-Dalsgaard}}{2012}]{jcd12}
{Christensen-Dalsgaard} J.,  2012, in {Shibahashi} H.,  {Takata} M.,
  {Lynas-Gray} A.~E.,  eds,  Astronomical Society of the Pacific Conference
  Series Vol. 462, Progress in Solar/Stellar Physics with Helio- and
  Asteroseismology. p.~503 (\mn@eprint {arXiv} {1110.5012})

\bibitem[\protect\citeauthoryear{{Christensen-Dalsgaard}
  et~al.,}{{Christensen-Dalsgaard} et~al.}{1996}]{jcd96}
{Christensen-Dalsgaard} J.,  et~al., 1996, \mn@doi [Science]
  {10.1126/science.272.5266.1286}, \href
  {http://adsabs.harvard.edu/abs/1996Sci...272.1286C} {272, 1286}

\bibitem[\protect\citeauthoryear{{Constantino}, {Campbell},
  {Christensen-Dalsgaard}, {Lattanzio}  \& {Stello}}{{Constantino}
  et~al.}{2015}]{constantino15}
{Constantino} T.,  {Campbell} S.~W.,  {Christensen-Dalsgaard} J.,  {Lattanzio}
  J.~C.,   {Stello} D.,  2015, \mn@doi [\mnras] {10.1093/mnras/stv1264}, \href
  {http://adsabs.harvard.edu/abs/2015MNRAS.452..123C} {452, 123}

\bibitem[\protect\citeauthoryear{{Cunha}, {Stello}, {Avelino},
  {Christensen-Dalsgaard}  \& {Townsend}}{{Cunha} et~al.}{2015}]{cunha15}
{Cunha} M.~S.,  {Stello} D.,  {Avelino} P.~P.,  {Christensen-Dalsgaard} J.,
  {Townsend} R.~H.~D.,  2015, \mn@doi [\apj] {10.1088/0004-637X/805/2/127},
  \href {http://adsabs.harvard.edu/abs/2015ApJ...805..127C} {805, 127}

\bibitem[\protect\citeauthoryear{{Deheuvels}, {Ballot}, {Beck}, {Mosser},
  {{\O}stensen}, {Garc{\'{\i}}a}  \& {Goupil}}{{Deheuvels}
  et~al.}{2015}]{deheuvels15}
{Deheuvels} S.,  {Ballot} J.,  {Beck} P.~G.,  {Mosser} B.,  {{\O}stensen} R.,
  {Garc{\'{\i}}a} R.~A.,   {Goupil} M.~J.,  2015, \mn@doi [\aap]
  {10.1051/0004-6361/201526449}, \href
  {http://adsabs.harvard.edu/abs/2015A%26A...580A..96D} {580, A96}

\bibitem[\protect\citeauthoryear{{Foreman-Mackey}, {Hogg}, {Lang}  \&
  {Goodman}}{{Foreman-Mackey} et~al.}{2013}]{emcee}
{Foreman-Mackey} D.,  {Hogg} D.~W.,  {Lang} D.,   {Goodman} J.,  2013, \mn@doi
  [PASP] {10.1086/670067}, 125, 306

\bibitem[\protect\citeauthoryear{{Gilliland} et~al.,}{{Gilliland}
  et~al.}{2010}]{gillilandetal10}
{Gilliland} R.~L.,  et~al., 2010, \mn@doi [\pasp] {10.1086/650399}, \href
  {http://adsabs.harvard.edu/abs/2010PASP..122..131G} {122, 131}

\bibitem[\protect\citeauthoryear{{Gough}}{{Gough}}{1993}]{gough93}
{Gough} D.~O.,  1993, in {Zahn} J.-P.,  {Zinn-Justin} J.,  eds, Astrophysical
  Fluid Dynamics - Les Houches 1987. pp 399--560

\bibitem[\protect\citeauthoryear{{Gough}}{{Gough}}{2002}]{Gough02}
{Gough} D.~O.,  2002, in {Battrick} B.,  {Favata} F.,  {Roxburgh} I.~W.,
  {Galadi} D.,  eds,  ESA Special Publication Vol. 485, Stellar Structure and
  Habitable Planet Finding. pp 65--73

\bibitem[\protect\citeauthoryear{{Gough}}{{Gough}}{2007}]{gough07}
{Gough} D.~O.,  2007, \mn@doi [Astronomische Nachrichten]
  {10.1002/asna.200610730}, \href
  {http://adsabs.harvard.edu/abs/2007AN....328..273G} {328, 273}

\bibitem[\protect\citeauthoryear{{Goupil}, {Mosser}, {Marques}, {Ouazzani},
  {Belkacem}, {Lebreton}  \& {Samadi}}{{Goupil} et~al.}{2013}]{goupil13}
{Goupil} M.~J.,  {Mosser} B.,  {Marques} J.~P.,  {Ouazzani} R.~M.,  {Belkacem}
  K.,  {Lebreton} Y.,   {Samadi} R.,  2013, \mn@doi [\aap]
  {10.1051/0004-6361/201220266}, \href
  {http://adsabs.harvard.edu/abs/2013A%26A...549A..75G} {549, A75}

\bibitem[\protect\citeauthoryear{{Hekker} \& {Christensen-Dalsgaard}}{{Hekker}
  \& {Christensen-Dalsgaard}}{2017}]{hekker17}
{Hekker} S.,  {Christensen-Dalsgaard} J.,  2017, \mn@doi [\aapr]
  {10.1007/s00159-017-0101-x}, 25, 1

\bibitem[\protect\citeauthoryear{{Hekker}, {Elsworth}  \& {Angelou}}{{Hekker}
  et~al.}{2018}]{hekker18}
{Hekker} S.,  {Elsworth} Y.,   {Angelou} G.~C.,  2018, \mn@doi [\aap]
  {10.1051/0004-6361/201731264}, \href
  {http://adsabs.harvard.edu/abs/2018A%26A...610A..80H} {610, A80}

\bibitem[\protect\citeauthoryear{{Houdek} \& {Gough}}{{Houdek} \&
  {Gough}}{2007}]{Houdek07}
{Houdek} G.,  {Gough} D.~O.,  2007, \mn@doi [\mnras]
  {10.1111/j.1365-2966.2006.11325.x}, \href
  {http://adsabs.harvard.edu/abs/2007MNRAS.375..861H} {375, 861}

\bibitem[\protect\citeauthoryear{{Jiang} \& {Christensen-Dalsgaard}}{{Jiang} \&
  {Christensen-Dalsgaard}}{2014}]{jiang14}
{Jiang} C.,  {Christensen-Dalsgaard} J.,  2014, \mn@doi [\mnras]
  {10.1093/mnras/stu1697}, \href
  {http://adsabs.harvard.edu/abs/2014MNRAS.444.3622J} {444, 3622}

\bibitem[\protect\citeauthoryear{{Kjeldsen} \& {Bedding}}{{Kjeldsen} \&
  {Bedding}}{1995}]{kjeldsen95}
{Kjeldsen} H.,  {Bedding} T.~R.,  1995, \aap, \href
  {http://adsabs.harvard.edu/abs/1995A%26A...293...87K} {293, 87}

\bibitem[\protect\citeauthoryear{{Miglio}, {Montalb{\'a}n}, {Noels}  \&
  {Eggenberger}}{{Miglio} et~al.}{2008}]{miglio08}
{Miglio} A.,  {Montalb{\'a}n} J.,  {Noels} A.,   {Eggenberger} P.,  2008,
  \mn@doi [\mnras] {10.1111/j.1365-2966.2008.13112.x}, \href
  {http://adsabs.harvard.edu/abs/2008MNRAS.386.1487M} {386, 1487}

\bibitem[\protect\citeauthoryear{{Monteiro}, {Christensen-Dalsgaard}  \&
  {Thompson}}{{Monteiro} et~al.}{1994}]{monteiroetal94}
{Monteiro} M.~J.~P.~F.~G.,  {Christensen-Dalsgaard} J.,   {Thompson} M.~J.,
  1994, A\&A, \href {http://adsabs.harvard.edu/abs/1994A%26A...283..247M} {283,
  247}

\bibitem[\protect\citeauthoryear{{Mosser} et~al.,}{{Mosser}
  et~al.}{2012}]{mosser12}
{Mosser} B.,  et~al., 2012, \mn@doi [\aap] {10.1051/0004-6361/201118519}, \href
  {http://adsabs.harvard.edu/abs/2012A%26A...540A.143M} {540, A143}

\bibitem[\protect\citeauthoryear{{Mosser}, {Vrard}, {Belkacem}, {Deheuvels}  \&
  {Goupil}}{{Mosser} et~al.}{2015}]{mosser15}
{Mosser} B.,  {Vrard} M.,  {Belkacem} K.,  {Deheuvels} S.,   {Goupil} M.~J.,
  2015, \mn@doi [\aap] {10.1051/0004-6361/201527075}, 584, A50

\bibitem[\protect\citeauthoryear{{Mosser}, {Gehan}, {Belkacem}, {Samadi},
  {Michel}  \& {Goupil}}{{Mosser} et~al.}{2018}]{mosser18}
{Mosser} B.,  {Gehan} C.,  {Belkacem} K.,  {Samadi} R.,  {Michel} E.,
  {Goupil} M.-J.,  2018, \mn@doi [\aap] {10.1051/0004-6361/201832777}, \href
  {http://adsabs.harvard.edu/abs/2018A%26A...618A.109M} {618, A109}

\bibitem[\protect\citeauthoryear{{Pedersen}, {Aerts}, {P{\'a}pics}  \&
  {Rogers}}{{Pedersen} et~al.}{2018}]{pedersen18}
{Pedersen} M.~G.,  {Aerts} C.,  {P{\'a}pics} P.~I.,   {Rogers} T.~M.,  2018,
  preprint, \href {http://adsabs.harvard.edu/abs/2018arXiv180202051P} {}
  (\mn@eprint {arXiv} {1802.02051})

\bibitem[\protect\citeauthoryear{{Shibahashi}}{{Shibahashi}}{1979}]{shibahashi79}
{Shibahashi} H.,  1979, \pasj, \href
  {http://adsabs.harvard.edu/abs/1979PASJ...31...87S} {31, 87}

\bibitem[\protect\citeauthoryear{{Takata}}{{Takata}}{2016}]{takata16}
{Takata} M.,  2016, \mn@doi [\pasj] {10.1093/pasj/psw104}, \href
  {http://adsabs.harvard.edu/abs/2016PASJ...68..109T} {68, 109}

\bibitem[\protect\citeauthoryear{{Tassoul}}{{Tassoul}}{1980}]{tassoul80}
{Tassoul} M.,  1980, \mn@doi [\apjs] {10.1086/190678}, \href
  {http://adsabs.harvard.edu/abs/1980ApJS...43..469T} {43, 469}

\bibitem[\protect\citeauthoryear{{Unno}, {Osaki}, {Ando}, {Saio}  \&
  {Shibahashi}}{{Unno} et~al.}{1989}]{unno89}
{Unno} W.,  {Osaki} Y.,  {Ando} H.,  {Saio} H.,   {Shibahashi} H.,  1989,
  {Nonradial oscillations of stars}

\bibitem[\protect\citeauthoryear{{Wu}, {Li}  \& {Deng}}{{Wu}
  et~al.}{2018}]{wu18}
{Wu} T.,  {Li} Y.,   {Deng} Z.-m.,  2018, \mn@doi [\apj]
  {10.3847/1538-4357/aadf85}, \href
  {http://adsabs.harvard.edu/abs/2018ApJ...867...47W} {867, 47}

\makeatother
\end{thebibliography}


\label{lastpage}

\clearpage

\appendix

\section{Signature of buoyancy glitches}
\label{apa}
The signature on the period spacing from a buoyancy glitch modelled by
a Dirac delta function has been derived by \cite{cunha15} based on the
variable  $\Psi=~(r^3/g\rho \tilde f) ^{1/2}\delta p$, where $\delta p$ is
the Lagrangian pressure perturbation and $\tilde f$ is a function of frequency and of the
equilibrium structure (the f-mode discriminant defined by equation
(35) of \cite{gough07}). Here we present
similar derivations for the cases of buoyancy glitches modelled by a step
function and by a Gaussian function, respectively. 

The starting point is the wave equation,
\begin{eqnarray}
\frac{\rmdd\Psi}{\rmd r^2}+K^2\Psi=0,
\label{waveeqap}
\end{eqnarray}
derived from the linear, adiabatic pulsation equations, for the case
of a spherically symmetric equilibrium
under the Cowling
approximation. The radial wavenumber $K$ is defined by,
\begin{eqnarray}
K^2=\frac{\omega^2-\omega_{\rm
    c}^2}{c^2}-\frac{L^2}{r^2}\left(1-\frac{\mN^2}{\omega^2}\right),
\label{k2}
\end{eqnarray}
where  $\omega_{\rm c}$ and $\mN$ are generalisations of the usual critical acoustic frequency and
buoyancy frequency, respectively, which account for all
terms resulting from the spherical geometry of the problem.  The exact forms of
these quantities can be found in equations (5.4.8) and (5.4.9) of
\cite{gough93}. In practice $\mN$ is very similar to $N$ throughout
the wave propagation cavity, where it will be relevant for our analysis, and, thus, we approximate the former by the latter
from the outset, similar to what has been done in
\cite{cunha15}. 

\subsection{Impact on pure gravity modes\label{apa1}}

In short, the derivation of the signature on the period spacing from the buoyancy
glitch is performed by considering the asymptotic solutions to
Eq.~(\ref{waveeqap}) on each side of the glitch and applying
appropriate matching conditions at the glich location. We recall that
the asymptotic solutions well inside the g-mode cavity, inwards and
outwards from the glitch position are, respectively 
\begin{equation}
\Psi_{\rm in} \sim \tilde\Psi_{\rm in} K_{\rm in}^{-1/2}
\sin\left(\int_{r_1}^r K_{\rm in}\rmd r +
  \frac{\pi}{4}\right),
\label{gasymp_l}
\end{equation}
and
\begin{equation}
\Psi_{\rm out} \sim \tilde\Psi_{\rm out} K_{\rm out}^{-1/2}
\sin\left(\int_r^{r_2} K_{\rm out}\rmd r +
  \frac{\pi}{4}\right),
\label{gasymp_r}
\end{equation}
where $\tilde\Psi_{\rm in}$ and $\tilde\Psi_{\rm out}$ are constants
and $K_{\rm in}$ and $K_{\rm out}$ refer to $K$ computed from $N_{\rm in}$ and
$N_{\rm out}$, respectively. 

\subsubsection*{Glitch modelled by a step function} 
In the case of the glitch modelled by the step-like function, the discontinuity in the buoyancy frequency at $r=r^\star$ leads to a
discontinuity in the wavenumber at the same position. Well inside the
g-mode cavity, $K\approx\frac{L N}{\omega \, r}$ and, thus, the relative amplitude of
the discontinuity in the wavenumber is given by,
\begin{equation}
\frac{\Delta K}{K_{\rm out}^\star} \approx \frac{N_{\rm in}^\star}{N_{\rm out}^\star}-1 = A_{\rm st},
\end{equation}
where $\Delta K =  \left. K_{\rm in}\right|_{ r\rightarrow r^\star_-}-\left. K_{\rm
  out}\right|_ {r\rightarrow r^\star_+}$ and the subscript $\star$
indicates that the quantities are to be taken at $r\rightarrow r_\pm^\star$. 

Similarly to the case of the glitch modelled by the Dirac delta function,
we impose the continuity of $\Psi$ at $r=r^\star$ \footnote{Strictly speaking,
  the continuity condition is satisfied by $\delta p$.  However, we
  have verified from the numerical solutions computed with \adipls\ that this condition is
  also very closely satisfied by $\Psi$.}. Moreover, by integrating
the wave equation (\ref{waveeqap}) once across the glitch,
letting the width of the region where the integration is performed tend to zero, it becomes evident
that the derivative of $\Psi$ must also be continuous at $r=r^\star$, unlike what was found in the case of the glitch modelled by the Dirac delta
function. This is because the integral of the step function is a
continuous function, while the integral of a Dirac delta function
is not. 

Imposing that both $\Psi$ and its derivative, taken asymptotically,
are continuous at the glitch position, we find,
\begin{eqnarray}
\sin\left(\int_{r_1}^{r_2}K\rmd 
  r+\frac{\pi}{2}\right)= && \nonumber\\
-A_{\rm st}
\sin\left(\int_{r^\star}^{r_2}K_{\rm out}\rmd
  r+\frac{\pi}{4}\right)\cos\left(\int_{r_1}^{r^\star}K_{\rm in}\rmd
  r+\frac{\pi}{4}\right).
\label{eigen_st}
\end{eqnarray}

Equation (\ref{eigen_st}) provides the eigenvalue condition in the
presence of a glitch modelled by a step-like function.  It differs in
two main aspects from the condition derived by \cite{cunha15} for the glitch modelled by
a Dirac delta function  (their equation 13). Firstly, the amplitude
multiplying the sinusoidal functions on the rhs is
independent of frequency, implying that the amplitude of the  signature of the glitch on
the period spacing will also be independent  of the
frequency in this case. Secondly, the rhs does not remain invariant when the
arguments inside the sinusoidal functions are interchanged,
highlighting the fact that in the present case glitches positioned symmetrically
about the center of the cavity produce different signatures. 

Next, we consider the specific case of a glitch located in the inner
half of the propagation cavity,  {\it i.e.} $\tilde\omega_{\rm g}^\star/\omega_{\rm g} < 0.5$ . Writing,
\begin{equation}
\int_{r^\star}^{r_2} K_{\rm out}\rmd r+\frac{\pi}{4} = \int_{r_1}^{r_2} K\rmd r+\frac{\pi}{2}-\int_{r_1}^{r^\star} K_{\rm in}\rmd r-\frac{\pi}{4},
\end{equation}
and substituting in Eq.~(\ref{eigen_st}) we find,
\begin{eqnarray}
\sin\left(\int_{r_1}^{r_2}K\rmd  r+\frac{\pi}{2}+\Phi\right) = 0,
\label{eigengap2}
\end{eqnarray}
where $\Phi$, and a new quantity, $B$, are defined by the following system of equations,
\begin{equation}
\left\{
\begin{array}{lll}
B\cos\Phi \hspace{-0.0cm}&\hspace{-0.0cm}= & \hspace{-0.0cm} 1+A_{\rm st}\cos^2\left(\int_{r_1}^{r^\star} K_{\rm in}\rmd r+\pi/4\right)
\\ \\
B\sin\Phi\hspace{-0.0cm}&\hspace{-0.0cm}= &\hspace{-0.0cm}
-\frac{1}{2}A_{~\rm st}\cos\left(2\int_{r_1}^{r^\star} K_{\rm in}\rmd r\right)
\end{array}
\right.
\label{Phi_st_in}
\end{equation}

Finally, to relate the phase $\Phi$ to the parameters characterising
the glitch, we approximate the integral in the arguments of the
sinusoidal functions in
Eq.~(\ref{Phi_st_in}) by $\int_{r_1}^{r^\star} K_{\rm in}\rmd r \approx \int_{r_1}^{r^\star}
\frac{L N}{\omega \, r} \rmd r +\delta \equiv \frac{\tilde\omega_{\rm g}^\star}{\omega}+\delta$.
This approximation follows from approximating $K$ by $\frac{LN}{\omega
\, r}$ inside the cavity. Because near the turning point, $r_1$, the wavenumber
approaches zero, this approximation leads to a slight overestimation of the value of the
wavenumber integral which is compensated by the introduction of the
phase $\delta$. The phase $\delta$ is, thus, related to the details of
mode reflection near the turning points of
the propagation cavity, more specifically, in the present case near
the inner turning point.  

The phase $\Phi$ defined by Eq.~(\ref{Phi_st_in}), with the
approximation described above, is then differentiated and used
in Eq.~(\ref{ps_coupling_glitch}) to derive the analytical expression
for the period spacing  given by Eq.~(\ref{ps_glitch_st_in}).

\subsubsection*{Glitch modelled by a Gaussian function} 

In the case of the glitch modelled by the Gaussian-like function, 
the variation in the buoyancy frequency around $r=r^\star$ produces a
variation in the wavenumber that can be expressed by 
	\begin{equation}
	\frac{\Delta K}{K_0}\approx\frac{A_{\rm G}}{\sqrt{2\pi}\Delta_{\rm g}}\exp{\left(-\frac{(\omega_{\rm g}^r-\omega_{\rm g}^{\star})^2}{2\Delta_{\rm g}^2}\right)},
	\end{equation}
	where, as before, we have assumed the glitch is located well inside
	the g-mode cavity and, thus, approximated the wave number by $K\approx\frac{L
		N}{\omega \, r}$, and defined the unperturbed wavenumber as $K_0\approx \frac{L
		N_0}{\omega \, r}$.

Similarly to what was done in previous cases, to establish the 
eigenvalue condition for this case, we need to
match the asymptotic solutions given by
Eqs~(\ref{gasymp_l})-(\ref{gasymp_r}) and their derivatives across
the glitch. We note, however, that unlike the cases of glitches modelled by a Dirac
delta function and by a step-like function, here the
glitch is not infinitely thin. Therefore, the exact matching would require
that we establish first how the eigenfunctions are perturbed inside 
the glich, which is unknown within the framework of our study,
because the solutions that we are employing were derived
asymptotically (hence, neglecting small-scale perturbations to the
background). 

{To proceed, we therefore make a significant simplification to the
problem which consists in assuming that the eigenfunction inside the glitch has the same functional form as that derived asymptotically in the absence of a glitch, with a slowly varying amplitude and a rapidly varying oscillatory part. In practice, this is achieved by extending the solutions on both sides of the glitch all the way to $r=r^\star$, keeping the amplitude proportional to the unperturbed $K_0^{-1/2}$.
Under this assumption, the continuity of $\Psi$ at $r=r^\star$ imposes that
\begin{equation}
\tilde\Psi_{\rm in} =\frac{\sin\left(\int_{r^\star}^{r_2} K\rmd r +
	\frac{\pi}{4}\right)}{\sin\left(\int_{r_1}^{r^\star} K\rmd r +
	\frac{\pi}{4}\right)}\tilde\Psi_{\rm out},
\label{contPsi}
\end{equation}
where $K_{\rm in}$ and $K_{\rm out}$ were assumed equal to $K$ on each side of $r^\star$, and the integration of the wave equation  (\ref{waveeqap}) across
the glitch gives,
\begin{equation}
\left[\frac{\rmd \Psi_{\rm out}}{\rmd r}-\frac{\rmd \Psi_{\rm
		in}}{\rmd r}\right]_{r^\star} = -\int_{r^\star-\Delta
	r}^{r^\star+\Delta r}\Delta K\Psi K\rmd r,
\label{contPsiprime}
\end{equation}
where $\pm\Delta r$ defines the region of impact of the glitch.}

{Equation~(\ref{contPsiprime}) shows that under our assumption the integrated impact of the glitch on the phase of the wave is taken at a single position, namely  $r=r^\star$. To compute the integral on the rhs of Eq.~(\ref{contPsiprime}) we need again to consider the eigenfunction inside the glitch. For mathematical consistency, we should take  the extended solutions on each side of $r^\star$. However, in the actual problem the phase does not jump at a single position. Thus, using the solution that incorporates that phase jump in the computation of the phase jump itself, is not necessarily a better approximation than adopting a solution that does not incorporate a phase  jump at $r^\star$, as would be achieved by taking the inner or the outer solution throughout the whole glitch.	 Given the above, we derive the eigenvalue conditions for both cases, and test their performance  aposteriori through the comparison with the numerical results. That comparison allows us also to check the implications of the simplification introduced in this analysis.}
	
{Adopting the extended solutions on each side of the glitch, and combining Eqs~(\ref{contPsi})-(\ref{contPsiprime}), we find, after some algebra, the eigenvalue condition,
\begin{eqnarray}
\scriptsize{\sin\left(\int_{r_1}^{r_2}K\rmd r+\frac{\pi}{2}\right)\left(1+\frac{A_{\rm G}\alpha}{\omega}\right)=} \nonumber\\
\scriptsize{A_{\rm G}f_{\omega}^{\Delta_{\rm g}}\sin\left(\int_{r_1}^{r_\star}K\rmd
	r+\frac{\pi}{4}\right)\sin\left(\int_{r_\star}^{r_2}K\rmd
	r+\frac{\pi}{4}\right)},
\label{eigen_G1}
\end{eqnarray}
where $\alpha=0.5\,\omega f_{\omega}^{\Delta_{\rm g}}\,{\rm erfi}(a)$, with $a=\sqrt{0.5}\, \Delta_{\rm g}\omega^{-1}$,  $f_{\omega}^{\Delta_{\rm g}}=\omega^{-1}{\rm
	e}^{{-a^2}}$, and $\rm erfi$ is the imaginary error function.}

{However, if we adopt either the extended inner solution or the extended outer solution throughout the whole integral, we find,
\begin{eqnarray}
\scriptsize{\sin\left(\int_{r_1}^{r_2}K\rmd r+\frac{\pi}{2}\right) =} \nonumber\\
\scriptsize{A_{\rm G}f_{\omega}^{\Delta_{\rm g}}\sin\left(\int_{r_1}^{r_\star}K\rmd
	r+\frac{\pi}{4}\right)\sin\left(\int_{r_\star}^{r_2}K\rmd
	r+\frac{\pi}{4}\right)}.
\label{eigen_G2}
\end{eqnarray}
Equation~(\ref{eigen_G1}) differs from Eq.~(\ref{eigen_G2}) due to the presence of the term $A_{\rm G}\alpha/\omega$ on the lhs. While $\alpha$ is always smaller than $\sim 0.3$, $A_{\rm G}/\omega$ may be large, as no assumption is made about the strength of the glitch. In that case, the two eigenvalue conditions will differ significantly. As we shall see, that difference will have an impact on the amplitude recovered when fitting the analytical period spacing derived for each case to the numerical one. }

{We note that both Eq.~(\ref{eigen_G1}) and (\ref{eigen_G2}) predict that the amplitude of the glitch signature is frequency dependent. This is unlike what was found for the step-like glitch ({\it cf.} Eq.~(\ref{eigen_st})). Moreover, in both cases we can note that the rhs remains invariant when the arguments inside the sinusoidal functions on the rhs are interchanged, highlighting that glitches modelled by a Gaussian function positioned symmetrically about the centre of the cavity produce similar signatures on pure gravity waves.  In reality, the requirement that the wave solutions are regular at the centre of the star is expected to introduce a slight asymmetry between the boundary conditions on the left and right sides of the g-mode cavity, leading to a slight asymmetry also in the glitch signature. That, however, is not accounted for in the asymptotic analysis presented  here and shall be subject to further discussion in future work. 
}

Next, we consider the specific case of a glitch located in the outer
half of the propagation cavity,  {\it i.e.} $\omega_{\rm g}^\star/\omega_{\rm g} < 0.5$ . Writing
\small
\begin{eqnarray}
\int_{r_1}^{r_\star}K\rmd r+\frac{\pi}{4}=\int_{r_1}^{r_2}K\rmd
r+\frac{\pi}{2}-\int_{r_\star}^{r_2}K\rmd r-\frac{\pi}{4}.
\label{int_out}
\end{eqnarray}
\normalsize
and substituting in Eqs~(\ref{eigen_G1}) and (\ref{eigen_G2}) we find
\begin{eqnarray}
\sin\left(\int_{r_1}^{r_2}K\rmd  r+\frac{\pi}{2}+\Phi\right) = 0,
\label{eigengap2}
\end{eqnarray}
where, $\Phi$ and $B$ take different forms, depending on the eigenvalue condition adopted. For the eigenvalue condition defined by (\ref{eigen_G1}) we find
\begin{equation}
\left\{
\begin{array}{lll}
B\cos\Phi \hspace{-0.0cm}&\hspace{-0.0cm}= & \hspace{-0.0cm}
1+\frac{A_{\rm G}\alpha}{\omega}-\frac{1}{2}{{A}_{\rm G}}f_{\omega}^{\Delta_{\rm g}}
\cos\left(2\int_{r_\star}^{r_2}K\rmd r
\right)
\\ \\
B\sin\Phi \hspace{-0.0cm}&\hspace{-0.0cm}= &\hspace{-0.0cm}
{{A}_{\rm G}}f_{\omega}^{\Delta_{\rm g}}\sin^2\left(\int_{r_\star}^{r_2}K\rmd
r+\frac{\pi}{4} \right)
\end{array}
\right.
\label{Phi_g1}
\end{equation}
while for the eigenvalue condition defined by (\ref{eigen_G1}) $\Phi$ and $B$ take the form
\begin{equation}
\left\{
\begin{array}{lll}
B\cos\Phi \hspace{-0.0cm}&\hspace{-0.0cm}= & \hspace{-0.0cm}
1-\frac{1}{2}{{A}_{\rm G}}f_{\omega}^{\Delta_{\rm g}}
  \cos\left(2\int_{r_\star}^{r_2}K\rmd r
  \right)
\\ \\
B\sin\Phi \hspace{-0.0cm}&\hspace{-0.0cm}= &\hspace{-0.0cm}
{{A}_{\rm G}}f_{\omega}^{\Delta_{\rm g}}\sin^2\left(\int_{r_\star}^{r_2}K\rmd
  r+\frac{\pi}{4} \right)
\end{array}
\right.
\label{Phi_g2}
\end{equation}

To relate $\Phi$ to the parameters characterising the glitch we
approximate $\int_{r_\star}^{r_2}K\rmd r
\approx \int_{r_\star}^{r_2}\frac{LN}{\omega r}\rmd r+\delta \equiv
\frac{\omega_{\rm g}^\star}{\omega}+\delta$, where in this case the phase $\delta$
is related to the details of the mode reflection near the outer turning
point. 

When the phase $\Phi$, with the approximation above, is differentiated and used in
Eq.~(\ref{ps_coupling_glitch}), the period spacing becomes
\begin{equation}
\frac{\Delta P}{\Delta P_{\rm as}}\approx\left[1-\mF_{\rm {G}}\right]^{-1},
\end{equation}
where $\mF_{\rm {G}}$ takes different forms, depending on whether we use Eqs~(\ref{Phi_g1}) or Eqs~(\ref{Phi_g2}).
If $\Phi$ and $B$ are defined by Eqs~(\ref{Phi_g1}), we find
\begin{eqnarray}
\mF_{\rm {G}}= -\hspace{-0.05cm}{\scriptsize
	\frac{A_{\rm G}
		f_{\omega}^{\Delta_{\rm g}}}{B^2}{\frac{\omega_{\rm g}^\star}{\omega_{\rm g}}}}\left\{ \left(1+\frac{A_{\rm G}\alpha}{\omega}\right)\cos\beta_1\right.\hspace{1.9cm} \nonumber && \\
\left.{\scriptsize+\left(-\frac{\omega}{\omega_{\rm g}^\star}\frac{\rmd \ln f_{\omega}^{\Delta_{\rm g}}}{\rmd\ln\omega}-
		A_{\rm G}
		f_{\omega}^{\Delta_{\rm g}}-\frac{A_{\rm G}a}{\sqrt\pi \omega_{\rm g}^\star}\right)\sin^2\beta_2}\right\}
\label{ps_glitch_g_ap1}
\end{eqnarray}
with $\beta_1=2\omega_{\rm g}^\star/\omega
+2\delta$, and
$\beta_2=\omega_{\rm g}^\star/\omega+\pi/4+\delta$. 
If, however, $\Phi$ and $B$ are defined by Eqs~(\ref{Phi_g2}), we find
\begin{eqnarray}
\mF_{\rm {G}}=\hspace{7cm} \nonumber && \\
-\hspace{-0.05cm}{\scriptsize
			\frac{A_{\rm G}
			f_{\omega}^{\Delta_{\rm g}}}{B^2}{\frac{\omega_{\rm g}^\star}{\omega_{\rm g}}\;\left[\cos\beta_1+\left(-\frac{\omega}{\omega_{\rm g}^\star}\frac{\rmd\ln f_{\omega}^{\Delta_{\rm g}}}{\rmd\ln\omega}-
			A_{\rm G}
			f_{\omega}^{\Delta_{\rm g}}\right)\sin^2\beta_2\right]}}
\label{ps_glitch_g_ap2}
\end{eqnarray}
Given the symmetry of the eigenvalue condition discussed before, the same expressions for the period spacing would be found if the glitch had been located in the inner half of the propagating cavity, but with $\omega_{\rm g}^\star$ replaced by $\tilde\omega_{\rm g}^\star$.

The analytical expressions provided by Eqs~(\ref{ps_glitch_g_ap1}) and (\ref{ps_glitch_g_ap2})
have been tested against the period spacing derived directly from
the numerical solutions to the pulsation equations in the absence of coupling
between p and g modes, which were computed with
ASTER, for our RGB-2 model. 
The results are shown in Table~\ref{tests}, where we confront the glitch parameters inferred from the fit of each expression to the numerical results and those measured directly from inspection of the buoyancy frequency. 

\begin{table*}
	\caption{{Second and third rows: parameters derived from the fit of the analytical expressions given by Eqs~(\ref{ps_glitch_g_ap1}) and (\ref{ps_glitch_g_ap2}) to the period spacing derived from ASTER for the RGB-2 model (at the
			luminosity bump).  Fourth and fifth rows: the same as the preceding rows, but with the function $f_{\omega}^{\Delta_{\rm g}}$  in the analytical expressions  replaced by $f_{\omega}^{\Delta_{\rm g}}=\omega^{-1}{\rm
				e}^{{-4a^2}}$ . The values shown correspond to the best fit model. For
			comparison, the values of the glitch parameters estimated directly
		from the buoyancy frequency obtained with ASTEC
	(Fig.~\ref{fig:glitch}\,d) are also shown on the last row. }}
	\label{tests} 
	\begin{minipage}{0.99\textwidth}
		\resizebox{\linewidth}{!}{%
			\begin{tabular}{|c|c|c|c|}
				\hline
				& $A_{\rm G}$ ($10^{-6}$ rad/s)  &
				$\omega_{\rm g}^\star$ ($10^{-6}$ rad/s)  & $\Delta_{\rm g}$  ($10^{-6}$ rad/s)  \\
				\hline
				Eq~(\ref{ps_glitch_g_ap1}) & 898 &  1763 &  255 \\
				\hline
				Eq~(\ref{ps_glitch_g_ap2}) & 602 & 1747 &  315 \\
				\hline
				Eq~(\ref{ps_glitch_g_ap1}); modified $f_{\omega}^{\Delta_{\rm g}}$  &803 & 1754 & 155  \\
				\hline
				Eq~(\ref{ps_glitch_g_ap2}); modified $f_{\omega}^{\Delta_{\rm g}}$  & 608 & 1748 &  159 \\
			   \hline
			  Estimated& 380 & 1632 & 156 \\
				\hline
			\end{tabular}
		}
	\end{minipage}
\end{table*}

While the analytical expressions given by Eqs~(\ref{ps_glitch_g_ap1}) and (\ref{ps_glitch_g_ap2}) provide a good fit to the numerical data (with a likelihood comparable to the fit illustrated in Fig.~\ref{caso1_best_g}), both the amplitude and the width of the glitch inferred from the fit are about twice the estimated value. 
To understand
the origin of that, we recall that the main impact expected from the width of the  Gaussian glitch is an attenuation of the glitch
signature with decreasing frequency, resulting from the fact that the local wavenumber approaches the characteristic scale of the glitch, as the frequency decreases. It, thus, seems quite likely that the
simplification introduced in the analysis presented here, namely,
accounting only for the integrated effect of the glitch on the phase
of the wave, is responsible for the differences seen in the numerical
and analytical period spacings.

Since the asymptotic approach adopted here precludes us from fully taking into
account the finite width of the glitch, it is important to compare the results from our simplified approach with those derived in the limit case of a small glitch, for which we can derive the impact of the glitch on the frequencies, to first order, without knowledge of the perturbed eigenfunctions.  We note that in this limit the eigenvalue condition expressed by Eq.~(\ref{eigen_G1}) approaches that given by  Eq.~(\ref{eigen_G2}) because $A/\omega$ is small. 
That analysis has, in fact, been carried out in previous works for the case
of a Gaussian-like acoustic glitch associated to the helium second
ionisation zone \citep{Gough02,Houdek07}. Those authors studied the
impact of that glitch on the p modes and found an
exponential decrease of the amplitude of the glitch signature with the
square of the frequency\footnote{We note that because the authors analysed the
  impact of an acoustic glitch on the p modes the dependence on
  frequency they found is, as expected, inverse to what is found
  here. While we find an exponential decrease with decreasing
  frequency squared, they find an exponential decrease with increasing
frequency squared.}, but with a factor of 4 greater than the one
found in the current analysis. Indeed, following their analysis, we recover the  analytical expression for the period spacing given by Eq.~(\ref{ps_glitch_g_ap2}) if the function $f_{\omega}^{\Delta_{\rm g}}$ is replaced by $f_{\omega}^{\Delta_{\rm g}}=\omega^{-1}{\rm
	e}^{{-4a^2}}$.

Motivated by that limit result, which should be satisfied by the more general expression,  we have
multiplied the exponent of the function $f_{\omega}^{\Delta_{\rm g}}$ by a factor of 4 and performed new fits. With this modification, the glitch width inferred from the fit of both analytical expressions to the numerical period spacings is brought into agreement with the value estimated directly from the buoyancy frequency, as seen from the two last rows of Table~\ref{tests}. The amplitudes, on the other hand, are hardly changed by the modification introduced. We, thus, find that both modified analytical expressions provide a good representation of the data (as illustrated in Fig.~\ref{caso1_best_g}, for the case of Eq.~(\ref{ps_glitch_g_ap2}) with the modified $f_{\omega}^{\Delta_{\rm g}}$), but both lead to an overestimation of the amplitude of the glitch.  We have tested the two modified expressions on an otherwise similar model, but with a glitch with an amplitude about three times larger. We found results very similar to those found for the model discussed here, where the position and width of the glitch are adequately recovered, but the amplitude is overestimated by the same factor as before ($\sim$ 1.5) in the case of Eq.~(\ref{ps_glitch_g_ap2}) and by a larger factor in the case of Eq.~(\ref{ps_glitch_g_ap1}). The fact that  Eq.~(\ref{ps_glitch_g_ap1}) is more complex than Eq.~(\ref{ps_glitch_g_ap2}) and results in a larger overestimation of the glitch amplitude, leads us to the conclusion that  Eq.~(\ref{ps_glitch_g_ap2}) with the modified  $f_{\omega}^{\Delta_{\rm g}}$  provides the best of the four options discussed here to fit the model data.


\subsection{Impact on mixed modes}

To combine the effect of mode coupling with that of a glitch modelled
by a Gaussian function, we follow again the analysis performed in
\cite{cunha15} for the Dirac-delta glitch.
When the waves propagate
also in the p-mode cavity, Eq.~(\ref{gasymp_r}) is substituted by {
\begin{equation}
\Psi_{\rm out} \sim \tilde\Psi_{\rm out} K_{\rm out}^{-1/2}
\sin\left(\int_r^{r_2} K_{\rm out}\rmd r +
  \frac{\pi}{4}+\varphi\right),
\label{gasymp_r_c}
\end{equation}
where, as for the case of the Gaussian-like glitch without
mode-coupling, this solution on the rhs of the glitch shall be extended all the way to $r=r^\star$, keeping the amplitude proportional to $K_0^{-1/2}$.} The frequency-dependent coupling phase $\varphi$ is defined by
Eq.~(\ref{varphi}) and expresses the
influence of the p-mode cavity on the wave solution.

For the specific case of a glitch located in the outer
half of the propagation cavity, as is our RGB-2 model, it follows that
the eigenvalue condition can be expressed as {
\begin{eqnarray}
\sin\left(\int_{r_1}^{r_2}K\rmd  r+\frac{\pi}{2}+\Phi+\varphi\right) = 0,
\label{eigengc}
\end{eqnarray}
where, following the conclusions of Sec.~\ref{apa1}}, now $\Phi$ and $B$ are defined by the following system of equations {
\begin{equation}
\left\{
\begin{array}{lll}
B\cos\Phi \hspace{-0.0cm}&\hspace{-0.0cm}= & \hspace{-0.0cm}
1-\frac{1}{2}{{A}_{\rm G}}f_{\omega}^{\Delta_{\rm g}}
  \cos\left(2\int_{r_\star}^{r_2}K\rmd r +2\varphi
  \right)
\\ \\
B\sin\Phi \hspace{-0.0cm}&\hspace{-0.0cm}= &\hspace{-0.0cm}
{{A}_{\rm G}}f_{\omega}^{\Delta_{\rm g}}\sin^2\left(\int_{r_\star}^{r_2}K\rmd
  r+\frac{\pi}{4}+\varphi\right),
\end{array}
\right.
\label{Phi_g_c}.
\end{equation}}
Here, as before, the function $f$ is formally derived to be {$f_{\omega}^{\Delta_{\rm g}}=\omega^{-1}{\rm
		e}^{{-a^2}}\equiv\omega^{-1}{\rm
  e}^{{-\frac{1}{2}\Delta_{\rm g}^2\omega^{-2}}}$}, but for the reasons
discussed  in Sections~\ref{g} and \ref{apa1} for the case of a
Gaussian-like glitch and no coupling, it will be replaced by  {$f_{\omega}^{\Delta_{\rm g}}=\omega^{-1}{\rm
  e}^{{-2\Delta_{\rm g}^2\omega^{-2}}}$} motivated by the low-amplitude glitch
limit and the numerical results.

Approximating, {$\int_{r_\star}^{r_2}K\rmd r
\approx \int_{r_\star}^{r_2}\frac{LN}{\omega r}\rmd r+\delta$} in Eqs~(\ref{Phi_g_c}),
differentiating $\Phi$, and substituting it in Eq.~(\ref{ps_coupling_glitch}), 
we finally find the period spacing given by Eqs~(\ref{F_c,g})-(\ref{b2gc}). 

Finally, we note that unlike in the case of the pure gravity modes, the signature on the mixed modes from the Gaussian-like glitch is not invariant to symmetric changes of the glitch about the center of the g-mode cavity. The reason is that the conditions on the left and right of the glitch are not the same, as a result of the p-mode cavity. Mathematically, that is seen from the comparison of Eq.~(\ref{gasymp_l}) with Eq.~(\ref{gasymp_r}) and Eq.~(\ref{gasymp_r_c}), respectively. While the first two are identical when considering equivalent buoyancy distances from each extreme of the cavity, the same is not true when the first and last of these equations are considered.

\section{Estimating the purely acoustic dipole-mode frequencies}
\label{apb}
In a red-giant star, the modes of degree $l=1$ have a mixed nature. Hence, their frequencies are different from those that pure acoustic modes would have in the same star.  Nevertheless, knowledge of those pure acoustic frequencies, $\omega_{{\rm a},n}$,  may be necessary to apply the analytic expressions provided in this work for the period spacing in the presence of mode coupling. One possible way to estimate those frequencies is to start from the frequencies of radial modes, which are also observed in red giants and are always purely acoustic. In Section \ref{c} we proposed two expressions to estimate the frequencies $\omega_{{\rm a},n}$. Here, we use the same expressions to estimate the frequencies of $l=1$ modes in a solar model. Because in the sun $l=1$ modes are purely acoustic, by doing so we can check the performance of each expression against the known purely acoustic model frequencies.

Inspired in the results of the asymptotic analysis \citep{tassoul80}, we write the frequencies of high radial-order purely acoustic modes in the following form,
\begin{equation}
\nu_{n,l}\approx \left(n+\frac{l}{2}\right)\Delta\nu_0-\frac{AL^2\Delta\nu_0^2}{\nu_{n,l}}+G\left(\nu_{n,l}\right),
\label{asymp}
\end{equation}
where $A$ is sensitive to the conditions in the innermost layers of the star and $G(\nu_{n,l})$ is a function of frequency that accounts for near surface effects considered, {\it e.g.}, in the asymptotic analysis by \cite{gough93}, and also for deviations from the asymptotically-derived frequencies introduced, {\it e.g.}, by the presence of acoustic glitches located inside the p-mode cavity. The important aspect to retain is that these frequency-dependent effects are present both when considering $l=0$ and $l=1$ modes. At the radial mode frequencies, the function $ G$ reduces to $G(\nu_{n,0})\approx\nu_{n,0}-n\Delta\nu_0$.

Considering that the function $G(\nu_{n,l})$ may vary slowly with frequency (meaning, on a scale of many radial orders), we can consider, first, the following rough approximation for  the frequencies of the $l=1$ modes,
\begin{equation}
\nu_{n,1}\approx\nu_{n,0}+\frac{1}{2}\Delta\nu_0+\frac{C}{\left(\nu_{n,0}+1/2\Delta\nu_0\right)},
\label{approx_1}
\end{equation}
where $C=-AL^2\Delta\nu_0^2$. Expressing the above in terms of angular frequencies, we find the option adopted in Eq.~(\ref{omega_a1}). 

Alternatively, we may try to account for the fact that, for a given
radial order, the function $G(\nu_{n,l})$ will take slightly different
values if considered at the frequency of the radial mode or at the
frequency of the dipole mode. That can in principle be done by
interpolating the function $G$ derived from the radial modes, at the
frequencies estimated for the dipole modes. If $G$ were indeed a slowly
varying function of frequency, one could consider fitting it
simultaneously across a number of $l=0$ radial orders prior to
interpolating it. However, one of the contributions to $G$ comes from the
acoustic glitches mentioned before, which may introduce significant
variations acccross just a few radial orders. For that reason, after
trying several fitting plus interpolation options we concluded that
the approach yielding the best results consists on linearly interpolating $G$ between each two consecutive radial mode frequencies taking its value at $\nu_{n,0}+1/2\Delta\nu_0$, which is the first-order estimate of the dipolar mode frequencies.   This option, applied to Eq.~(\ref{asymp}), leads us to the following estimate of the dipole-mode acoustic frequencies:
\begin{equation}
\nu_{n,1}\approx\left(n+\frac{1}{2}\right)\Delta\nu_0+\frac{C}{\left(\nu_{n,0}+1/2\Delta\nu_0\right)}+G\left(\nu_{n,0}+1/2\Delta\nu_0\right),
\label{approx_2}
\end{equation}
where the last term on the rhs is to be interpreted as the value of the function $G$ obtained from interpolation at the frequencies defined by the expression within the brackets. This estimate, expressed in terms of angular frequencies,  provides the option adopted in Eq.~(\ref{omega_a2}).

\begin{figure} 
         \begin{minipage}{1.\linewidth}  
               \rotatebox{0}{\includegraphics[width=1.0\linewidth]{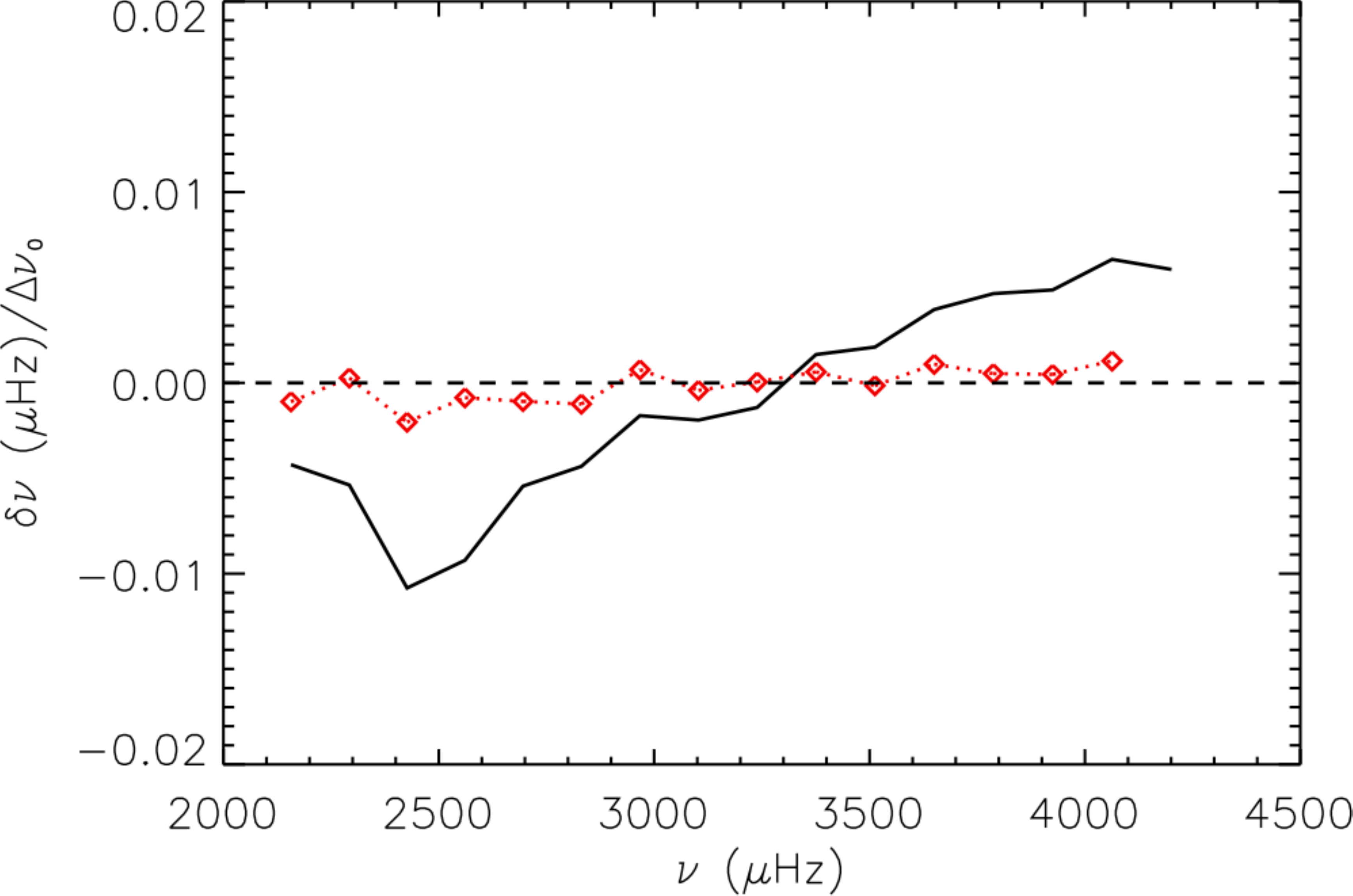}}
         \end{minipage}
\caption{Difference between the $l=1$ acoustic frequencies, $\nu_{n,1}$, computed with ADIPLS for the
  solar model S \citep{jcd96} and those estimated from: i) Eq.~(\ref{approx_1})
  (grey curve) and ii) Eq.~(\ref{approx_2}) (red-dashed curve;
  diamonds). The difference is scaled by the average solar large frequency separation computed from the $l=0$ modes within the range of radial orders shown, namely $13\le n\le 28$. In this case the constant $C = -AL^2\Delta\nu_0^2$ was estimated through uniformly weighted averages over the same range of $n$,
  $\left< (\nu_{n,1}-\nu_{n,0}-0.5\Delta\nu_0)(\nu_{n,0}+0.5\Delta\nu_0)\right>$ and $\left<(\nu_{n,1} -
(n+0.5)\Delta\nu_0-G(\nu_{n,0}+0.5\Delta\nu_0))(\nu_{n,0}+0.5\Delta\nu_0)\right>$,
respectively.
}
   \label{comp_solar}
       \end{figure}

The comparison between the model S dipole-mode frequencies, computed with ADIPLS, and the estimates proposed by Eqs~(\ref{approx_1}) and (\ref{approx_2}), corresponding to the options 1. and 2., respectively, in Sec.~\ref{c}, is shown in Fig~\ref{comp_solar}.  As expected, Eq.~(\ref{approx_2}) (option 2. in Sec.~\ref{c}) represents more closely the true model S dipole-mode frequencies. The differences that are still found when that option is considered stem from the fact that acoustic glitches introduce frequency variations that are not fully accounted for by the linear interpolation between radial modes considered here. 

\bsp	
\label{lastpage}
\end{document}